\documentclass[prd,superscriptaddress,altaffilletter,showpacs,nofootinbib]{revtex4}
\usepackage[dvips]{graphicx}
\usepackage{amsmath}

\newcommand{\be}{\begin{equation}}
\newcommand{\ee}{\end{equation}}
\newcommand{\bea}{\begin{eqnarray}}
\newcommand{\eea}{\end{eqnarray}}

\begin{document}

\title{Global asymptotic dynamics of the cubic galileon interacting with dark matter}


\author{Roberto De Arcia}\email{rc.dearcia@ugto.mx}\affiliation{Departamento Ingenier\'ia Civil, Divisi\'on de Ingenier\'ia, Universidad de Guanajuato, C.P. 36000, Gto., M\'exico.}

\author{Israel Quiros}\email{iquiros@fisica.ugto.mx}\affiliation{Departamento Ingenier\'ia Civil, Divisi\'on de Ingenier\'ia, Universidad de Guanajuato, C.P. 36000, Gto., M\'exico.}

\author{Ulises Nucamendi}\email{ulises.nucamendi@umich.mx}\affiliation{Instituto de F\'isica y Matem\'aticas, Universidad Michoacana de San Nicol\'as de Hidalgo, Edificio C-3, Cd Universitaria, C.P. 58040, Morelia, Michoac\'an, M\'exico.}

\author{Tame Gonzalez}\email{tamegc@ugto.mx}\affiliation{Departamento Ingenier\'ia Civil, Divisi\'on de Ingenier\'ia, Universidad de Guanajuato, C.P. 36000, Gto., M\'exico.}


\begin{abstract} In this paper we perform a thorough dynamical systems analysis of the cubic galileon model non-minimally coupled to the dark matter. Three well-known classes of interacting models are considered where the energy exchange between the dark components  is a function of the dark matter density and of the dark energy density: $Q_1=3\alpha H\rho_m$, $Q_2=3\beta\rho_m\dot{\phi}$ and $Q_3=3 \epsilon H  \rho_\phi$, respectively. We are able to show the global asymptotic dynamics of the model for the exponential potential in a homogeneous and isotropic background. The cosmological implications of the proposed scenarios are explored and it is found that, in addition to the appearance of new equilibrium configurations that do not appear neither in the non-interacting cubic galileon model nor in the interacting quintessence model, there is a significant impact of the non-minimal coupling through modification of the stability properties of the critical points. The resulting cosmological scenario provides a bigbang origin of the cosmic expansion, an early transient stage of inflationary expansion, as well as matter-scaling late time stable state of the universe, among other solutions of lesser cosmological interest. This work extends previous studies of coupled dark energy to a broader class of gravitational theories.\end{abstract}


\maketitle

\date{\today}


\section{Introduction}\label{sec-intro}

An increasing number of astronomical and cosmological observations such as studies of the luminosity of SNe Ia, CMB temperature anisotropies  and the measurement of Baryon Acoustic Oscillations, to name a few, indicate that our universe is undergoing a phase of accelerated expansion \cite{Riess1998,Perlmutter1999,Betoule2014,Anderson2014,Story2013,Hou2014,Giannantonio2014,Bartelmann2001,Cabass2015,Fernandez2014,Akrami2020}. If one assumes General Relativity (GR) as the theory that correctly describes the gravitational interactions at cosmological scales, these observations require to introduce new components: the ``dark matter''  (DM) and the ``dark energy'' (DE). The former is required in order to meet the cosmological observations on structure formation, while the latter is necessary to get late time acceleration of the expansion. The simplest model of dark energy included in the Lambda Cold Dark Matter ($\Lambda$CDM) cosmology, assumes the existence of a non vanishing cosmological constant satisfying the equation of state (EoS) $p=-\rho$, which is usually interpreted as the quantum vacuum energy. Although it is the simplest model that provides a reasonably account with observations, it is plagued by some theoretical and philosophical problems (at both the classical and quantum level) such as the vacuum energy problem \cite{Weinberg1989,Carroll2001} and the cosmic coincidence problem \cite{Zlatev1999}. Moreover, as the accuracy of cosmological observations increases, tensions among different data sets have also emerged and this might be the first sign for physics beyond the $\Lambda$CDM model. For reviews see \cite{Verde2019} and \cite{DiValentino2021}.

Light interacting scalar field models have been proposed as viable candidates for dynamical dark energy \cite{Ratra1998,Caldwell1998,Wang2000}. Quintessence is the simplest canonical scalar field and it is characterized by its EoS: $\omega_\phi=p_\phi/\rho_\phi$ ($-1\leq\omega_\phi<-1/3$), where $p_\phi$ and $\rho_\phi$ denote the pressure and the energy density, respectively. If the accelerating phase is an attractor solution that is independent of the initial conditions, then the coincidence problem can be solved, or at least alleviated at some level \cite{Amendola2000,Pavon2001}. Phantom fields were first introduced in cosmology as toy models in which the equation of state takes values below the ``phantom divide'': $\omega=-1$. In this scenario, the kinetic term has a negative sign  which implies a violation of the null energy condition and then these models are known to face the problem of future curvature singularity \cite{Caldwell2002,Carroll2003}. More general approaches are based on postulation of non-standard kinetic terms in the scalar field Lagrangian leading to the so-called ``k-essence'' models. Initially conceived to describe the inflationary universe, the k-essence theories have also been used to describe the present phase of accelerated expansion as driven by the kinetic term instead of the self-interacting potential \cite{Armendariz2001}. In a cosmological context, one of the characteristics of these theories is that under some hypothesis these can reproduce a fluid dynamics at background and perturbative levels in classical and quantum scenarios \cite{Bronnikov2016,Kleidis2018}.

On the other hand, the variety of modified theories of gravity that have been explored in the search for answers to open questions related to the present inflationary stage of the cosmic evolution, is impressive. Among these theories are included the Brans-Dicke (BD) theory, scalar-tensor theories (STT), the $f(R)$ theory and theories with higher-curvature modifications \cite{Brans1961,Nordtvedt1970,Wagoner1970,Damour1993,Damour19932,Damour1996,Nojiri2011,Sotiriou2006,Arciniega2020} (see also \cite{DeFelice2010,Quiros2019r,Cliftonr,Copelandr} for reviews). In the framework of BD theory the carriers of the gravitational interactions are the two polarizations of the graviton and the scalar field. STT-s are a generalization of the BD theory where the coupling parameter is a function of the scalar field: $\omega_\text{BD}\rightarrow\omega(\phi)$, i. e., it is a varying parameter. The need for a generalization of BD theory comes from the tight constraints on $\omega_\text{BD}$ that the solar system experiments have established \cite{Will}, although these constraints are smoothed out by the chameleon mechanism in BD theories with a self-interacting scalar field \cite{Khoury2004,Brax2004,Quiros2015}. If $\omega_\text{BD}$ were a varying coupling then the latter experimental constraints may be avoided or, at least, alleviated. It is well-known that under certain conditions the $f(R)$ theories are mathematically equivalent to the BD theory.

Further generalization of scalar-tensor theories exhibiting second order equations of motion in 4 dimensions are known as ``galileon''  and also as Horndeski theories \cite{Nicolis2009,Horndeski1974,Deffayet2009,Deffayet20092}. In terms of galileon models, the action can be expressed through linear combinations of the following Lagrangians:

\begin{align}
&{\cal L}_2=K(\phi,X),\nonumber \\
&{\cal L}_3 =-G_3(\phi,X)(\nabla^2 \phi),\nonumber\\
&{\cal L}_4=G_4(\phi,X)R + G_{4,X}\left[(\nabla^2 \phi)^2-2(\nabla_\mu \phi \nabla_\nu \phi)^2\right],\nonumber\\
&{\cal L}_5=G_5(\phi,X) G_{\mu\nu}\nabla^{\mu}\nabla^\nu\phi-\frac{1}{6}G_{5,X}\left[(\nabla^2 \phi)^3-3  (\nabla^2 \phi)(\nabla_\mu\nabla_\nu\phi)^2+2(\nabla_\mu\nabla_\nu\phi)^3\right].\nonumber\end{align} The functions $K=K(\phi,X)$ and $G_i=G_i(\phi,X)$ $(i=3,4,5)$ depend on the scalar field $\phi$ and on its kinetic energy density $X=-(\nabla \phi)^2/2$, while  $G_{i,X}$ represents the derivative of the functions $G_i$ with respect to  $X$. As usual, the Einstein tensor is given by $G_{\mu\nu}=R_{\mu\nu}-g_{\mu \nu}R/2$, $(\nabla\phi)^2:=g^{\mu\nu}\nabla_\mu\phi \nabla_\nu\phi$ and the D'Alembertian operator reads $\nabla^2 \phi:= g^{\mu\nu} \nabla_\mu\nabla_\nu\phi$. Finally, $(\nabla_\mu \phi \nabla_\nu \phi)^2 := (\nabla_\mu \phi \nabla_\nu \phi) (\nabla^\mu \phi \nabla^\nu \phi)$ and $(\nabla_\mu \phi \nabla_\nu \phi)^3 := (\nabla_\mu \phi \nabla^\beta \phi) (\nabla_\beta \phi \nabla^\alpha \phi) (\nabla_\alpha \phi \nabla_\nu \phi).$

A particular  class of cubic galileon model is depicted by
\be K=X-V,\,\,\, G_3=\sigma X, \,\,\, G_4=\frac{1}{16\pi G_N}, \,\,\, G_5=0, \label{model}\ee where $V=V(\phi)$ is the self-interacting potential of the scalar field, $\sigma$ is a non-negative coupling parameter and $G_N$ is the Newton's constant. This scenario survives cosmological observational tests, in particular the one related with the nearly simultaneous detection of gravitational waves GW 170817 \cite{Abbott2017} and the $\gamma$-ray burst GRB 170817A \cite{Abbott20172}. For the choice \eqref{model} the resulting action reads:

\bea S=\int d^4x\frac{\sqrt{-g}}{2}\left\{R-\left[1+\sigma\,\nabla^2\phi\right](\nabla\phi)^2 - 2 V +2{\cal L}_m \right\},\label{action}\eea with ${\cal L}_m={\cal L}_m(g_{\mu\nu},\psi)$ being the Lagrangian of the matter degrees of freedom $\psi$. This scenario has been proved to be interesting from  theoretical considerations and it is in good agreement with recent astrophysical and cosmological observations \cite{Deffayet2010,Babichev2013}. In \cite{Leon2013,Dearcia2016,Dearcia2018} the authors performed a dynamical system analysis of a non interacting cubic galileon model with dark matter as the source of gravity and for the vacuum as well, for a wide range of self-interacting potentials. Appreciable modification of the late-time dynamics was found only for the vacuum theory. In the presence of background matter the late-time dynamics is not modified in any appreciable way by the galileon.

The several of the above mentioned issues could be alleviated, in principle, if we assume that the DM and the DE do not evolve separately, but interact non-gravitationally \cite{Amendola1999,Amendola2000,Pavon2001,Coley2000,Chimento2003,Bohmer2008}. Although these models were first introduced in order to justify the currently small value of the cosmological constant, these were found to be very useful in order to
alleviate the coincidence problem, as well as to improve the predictions of the $\Lambda$CDM model on structure formation, BAO, CMB anisotropies, galaxy clusters and $H(z)$ data \cite{Cai2005,Salvatelli2014,Guendelman2016,Divalentino2020} (see also the reviews \cite{Bolotin2015,Wang2016} and references therein).

Non-minimal coupling can play an effective role to alleviate/solve the Hubble constant tension \cite{DiValentino2021,Divalentino2020}. Several interacting models significantly increase the $H_0$ value. The simplest interacting scenario is the case where vacuum energy with EoS 
\bea \omega_\text{DE}=\frac{p_\text{DE}}{\rho_\text{DE}}=-1,\nonumber\eea interacts with DM. This is known as the interacting vacuum scenario \cite{DiValentino2021,634,635,636,638,639,640,641,642}. Observational evidence shows that these models can solve the $H_0$ tension in an exceptional way \cite{DiValentino2021}. For instance, if consider the coupling term $Q=\alpha H\rho_m$, the Hubble tension may be completely solved within $1\sigma$ confidence \cite{639}. There are other models with non-gravitational interactions, such as the coupled scalar field, which either completely solves the problem or alleviates it with variable confidence levels.

Because of our current lack of knowledge about the nature of the two dark components (DM and DE) of the cosmic budget, it would be imprudent to discard additional non gravitational interactions between them.\footnote{Given that dark matter and dark energy do not interact with radiation and interact with matter only very weakly, assuming an additional non-gravitational interaction between these predominant components of the cosmic budget does not contradict the stringent constraints coming from Solar system experiments on fifth-force.} A wide variety of phenomenological forms of the interacting term at the level of the cosmological field equations have been proposed giving rise to a richer cosmological background dynamic compared with non-interacting models \cite{Gannouji2012}. Besides, theoretical aspects such as the possibility of constructing a Lagrangian from which the interaction term can be derived through using Lagrangian multipliers, have been studied in \cite{Bohmer2015}. The application of the pullback formalism for fluids has been analysed in \cite{Pourtsidou2013}. In \cite{Gonzalez2018} the authors constraint the EoS parameter of the DE component non-minimally coupled with DM in light of the second law of thermodynamics and the positiveness of entropy. The consequences of the energy conditions in order to construct reasonable models have also been considered in \cite{Huey2006,Folomeev2014}. Current cosmological data are compatible with such energy transfer. However the evidence so far is not conclusive \cite{Costa2014}.

In this work, we only consider particular linear interacting couplings in the sense defined by \cite{Chimento2010,Amendola2007,Pavon2009,Quartin2008}, however, inside of the general relativity framework, nonlinear interacting couplings have also been studied in the literature \cite{Chimento2010,Mangano2003,Lip2011,Baldi2011,Arevalo2012,Yang2018,Paliathanasis2019}. Let us mention that, less studied, some specific nonlocal interacting couplings have been analyzed, by example, in \cite{Amendola2008} a nonlocal dilaton coupling to dark matter was studied and shown that the nonlocal effects generate a backreaction which, for strong coupling, can automatically compensate the acceleration due to the negative pressure of the dilaton potential, thus asymptotically restoring the standard decelerated regime dominated by dust.

In this paper we want to generalise former works by allowing for an additional non-gravitational interaction between the cubic galileon and the dark matter exclusively. Recall that additional non-gravitational interactions with standard matter (baryons, radiation, etc.) are strongly constrained by Solar System experiments \cite{Will}. Our aim is to perform a thorough analysis of the global asymptotic dynamics of the mentioned interacting scenario, so that we could compare our results with the results of previous dynamical systems studies of the cubic galileon model without the additional interaction \cite{Leon2013,Dearcia2016,Dearcia2018}, and of the coupled quintessence \cite{Coley2000,Chimento2003,Bohmer2008}. These are particular cases of the present setup when both, the additional non-gravitational interaction and the cubic interaction, are turned off. 

The organization of this work is as follows. In section \ref{basic}, we expose the fundamentals of the interacting cubic galileon in the cosmological framework and we introduce helpful physical quantities. In section \ref{dynan}, through using suitable dimensionless dynamical system's variables, the cosmological field equations are traded by an equivalent dynamical system that is defined in some phase space. We focus on three well-known phenomenological expressions for the interacting term $Q$, namely $Q_1=3\alpha H\rho_m$ \cite{Coley2000,Bohmer2008,Wang2005,Riess2016,Aghanim2016,Li2020,Mifsud2019,Baldi2011,Li2019}, $Q_2=3\beta\rho_m\dot{\phi}$ \cite{Coley2000,Bohmer2008,Zhang2005,Tocchini2002,Costa2015,Cicoli2012,Barros2019} and $Q_3= 3 \epsilon H \rho_\phi$ \cite{Pavon2008,Yang2018,Pan2019,Yang2020,Yang2021,Kundu2021,Cardenas2019}. The corresponding dynamical systems are investigated and the fixed points, as well as their existence and stability properties, are determined. In section \ref{sec-discuss} we discuss on the cosmological implications and on the physical behaviour of the scenarios. Finally, in section \ref{sec-conclu}, we summarise the most relevant results.


\section{Cosmological equations} \label{basic}

For a spatially flat Friedmann-Lema\^{i}tre-Robertson-Walker (FLRW) spacetime background metric with line element $ds^2=-dt^2+a^2(t)\delta_{ik}dx^i dx^k$, the cosmological equations of motion that are derived from the action Eq. \eqref{action}, read:

\bea &&3H^2=\rho_m+\rho_\phi, \label{ef1} \\
&&-2\dot H=\rho_m + p_m+\rho_\phi+p_\phi,\label{ef2}\\
&&\dot{\rho}_m+3H(\rho_m+p_m)=Q,\label{ecm}\\
&&\left(1-6\sigma H\dot\phi\right)\dot\phi\ddot\phi+3H\dot\phi^2-3\sigma(\dot H+3 H^2)\dot\phi^3=-V_{,\phi}\dot{\phi}-Q,\label{ecc}\eea where $\rho_m$ and $p_m$ denote the energy density and the pressure of the matter fluid, respectively. The  interaction term $Q$ represents the transfer of energy in the dark sector: for $Q>0$ the energy flows from the galileon to the DM and the contrary for $Q<0$. It is important to mention that the equations \eqref{ecm}-\eqref{ecc} are not independent of each other due to the Bianchi identities. To proceed further we define the effective energy density and pressure:

\bea \rho_\phi=\frac{\dot\phi^2}{2}\left(1-6\sigma H\dot\phi\right)+V(\phi),\hspace{1cm} p_\phi=\frac{\dot\phi^2}{2}\left(1+2\sigma\ddot\phi\right)-V(\phi),\label{rho-p}\eea respectively, which are related through the effective dark energy equation of state (EoS): $\omega_\phi=p_\phi/\rho_\phi$. Notice that in the formal limit $\sigma\rightarrow 0$, the interacting quintessence scenario of Refs. \cite{Amendola2000,Pavon2001} is recovered from the present model, while in the limit $Q\rightarrow 0$, the non-interacting cubic galileon model of Refs. \cite{Leon2013,Dearcia2016,Dearcia2018} is approached. 

It will be useful to write Eqs. \eqref{ef2} and \eqref{ecc} in the following form,

\bea \frac{\dot H}{H^2}=-\frac{3}{2}\left(1+\omega_m\Omega_m\right)-\frac{\dot\phi^2}{4H^2}+\frac{V}{2H^2}-\frac{\sigma\dot\phi^2\ddot\phi}{2H^2},\label{usef-1}\eea and

\bea \frac{\ddot\phi}{H^2}=\frac{-3\frac{\dot\phi}{H}-\frac{V_{,\phi}}{H^2}-\frac{Q}{H^2\dot\phi}+3\sigma\dot\phi^2\left[\frac{3}{2}(1-\omega_m\Omega_m)-\frac{\dot\phi^2}{4H^2}+\frac{V}{2H^2}\right]}{1-6\sigma H\dot\phi+\frac{3\sigma^2\dot\phi^4}{2}},\label{usef-2}\eea respectively.


\section{Dynamical system analysis} \label{dynan}

Our aim will be to trade the system of second-order equations \eqref{ef1}-\eqref{ecc}, by a system of autonomous ODEs of the form $\dot{\vec{x}}=\vec{f}(\vec{x})$. Here $\vec{x}$ is the vector of a point in the phase space and the overdot means derivative with respect to the time variable. Since $\vec{f}$ does not depend explicitly on the time variable, the corresponding ODE system is autonomous. A critical point $\vec{x}_c$ satisfies the condition $\vec{f}(\vec{x}_c)=\vec{0}$. In order to determine the stability of the critical point we expand around $\vec{x}_c$ as $\vec{x}=\vec{x_c}+\vec{u}$, with $\vec{u}$ the vector of the perturbations. Therefore, for each critical point we obtain the equations for the perturbation up to the first order: $\dot{\vec{u}}= I\!\!M \vec{u}$, where $I\!\!M$ is the linearization matrix.\footnote {Thanks to the Hartman-Grobman theorem \cite{hartman,grobman}, which basically states that the behavior of a dynamical system in the neighbourhood of each hyperbolic fixed point is qualitatively the same as the behavior of its linearization, we can safely replace the study of the dynamics of the original dynamical system by the corresponding study of its linearization.} It is to be evaluated at each critical point in order to determine the stability through the analysis of the corresponding eigenvalues of $I\!\!M(\vec{x_c})$. In other words, thanks to the Hartman-Grobmann theorem, we trade the original dynamical system by the linearization around its hyperbolic solutions. The critical points in the phase space will correspond to relevant solutions of the cosmological equations.

In order to write \eqref{ef1}-\eqref{ecc} in the form of an autonomous dynamical system we use an appropriate set of normalized and bounded variables \cite{Dearcia2016}:

\bea
x_\pm=\frac{1}{x_s\pm1},\,\,\,\,\,\,\,y=\frac{1}{y_s+1},\,\,\,\,\,\,\, z=\frac{1}{H^2 \sigma +1},\label{n-var}
\eea where $x_+$ is for increasing $\phi$-s ($\dot\phi\geq 0$) and $ x_- $ is for decreasing $\phi$-s ($\dot\phi\leq 0$), besides, $0 \leq x_+ \leq 1 \,(-1 \leq x_- \leq 0), \, 0 \leq y \leq 1, \, 0 \leq z \leq 1$. Here $x_s$ and $ y_s$ are the Hubble normalized variables introduced in \cite{cope}: 

\bea x_s=\frac{\dot\phi}{\sqrt{6}H},\;y_s=\frac{\sqrt V}{\sqrt{3}H},\label{normal-vars}\eea and we are assuming expanding universes exclusively: $H\geq 0$. In this paper, for definiteness, we choose an exponential self-interacting potential:
 
\bea V(\phi)=V_0\,e^{-\lambda\phi},\label{exp-pot}\eea where $V_0$ and $\lambda$ are positive constants. Models with these characteristics have been the subject of  interest and arise naturally from theories of gravity such as scalar-tensor theories and string theories \cite{pot}. The cosmological equations \eqref{ef1}-\eqref{ecc} can be written in the equivalent form of the following dynamical system:

\begin{align} &x'_\pm =-\frac{x_{\pm}^2}{\sqrt{6}}\left(\frac{\ddot{\phi}}{H^2}\right)_\pm+x_{\pm}(1\mp x_{\pm})\left(\frac{\dot{H}}{H^2}\right)_\pm,\label{ode1}\\
&y'=y(1-y)\left[\sqrt{\frac{3}{2}}\lambda\left(\frac{1\mp x_{\pm}}{x_{\pm}}\right)+\left(\frac{\dot{H}}{H^2}\right)_\pm\right],\label{ode2}\\
&z'=-2z(1-z)\left(\frac{\dot{H}}{H^2} \right)_\pm,\label{ode3}\end{align} where the prime denotes derivative with respect to the number of $e$-foldings: $N=\ln a$ ($a$ is the scale factor). 

In terms of bounded variables \eqref{n-var} for Eqs. \eqref{usef-1} and \eqref{usef-2} we have that,

\begin{align}
\left(\frac{\dot{H}}{H^2}\right)_\pm =&-\frac{3}{2}(1+\omega_m\Omega_m)-\frac{3}{2}\left(\frac{1 \mp x_\pm}{x_\pm}\right)^2+\frac{3}{2}\left(\frac{1-y}{y}\right)^2-3\left(\frac{1 \mp x_\pm}{x_\pm}\right)^2\left(\frac{1-z}{z}\right)\left(\frac{\ddot\phi}{H^2}\right)_\pm,\label{dotH}\\
\left(\frac{\ddot\phi}{H^2}\right)_\pm =&\frac{27(1\mp x_\pm)^2z(1-z)}{y^2D_\pm}\left[x^2_\pm y^2(1-\omega_m\Omega_m)-(1\mp x_\pm)^2y^2+x^2_\pm(1-y)^2\right]\nonumber\\
&-\frac{x^3_\pm z^2}{\sqrt{6} (1\mp x_\pm)y^2D_\pm}\left[18(1\mp x_\pm)^2y^2-3\sqrt{6}\lambda x_\pm(1\mp x_\pm)(1-y)^2-x^2_\pm y^2\frac{Q}{H^3}\right],\label{ddotphi}\end{align} where we have defined

\bea D_{\pm}:=x_\pm^4 z^2-6\sqrt{6}x_\pm^3(1\mp x_\pm)z(1-z)+54(1 \mp x_{\pm})^4(1-z)^2.\nonumber\eea Here we shall consider non-relativistic (dust-like) perfect fluid describing dark matter $(p_m=0)$, i. e., $\omega_m=0$.

The Friedmann constraint \eqref{ef1} reads:

\bea \Omega_m=1-\frac{(1 \mp x_\pm)^2}{x_\pm^2}-\frac{(1-y)^2}{y^2}+\frac{6\sqrt{6}(1\mp x_\pm)^3(1-z)}{x_\pm^3 z},\label{fried-const}\eea and it is used to eliminate $\Omega_m$ from the equations. The bounded phase space where to look for the critical points of the dynamical system \eqref{ode1}-\eqref{ode3} is given by:

\bea \Psi=\Psi^+\cup\Psi^-,\;\Psi^\pm=\left\{(x_\pm,y,z):\,|x_\pm|\leq 1,\,\,0\leq y\leq 1,\,\,z\leq z^\pm_\text{surf}(x_\pm,y)\right\},\label{psi}\eea where the following bounding surfaces have been defined:

\bea z^\pm_\text{surf}(x_\pm,y)=\frac{6\sqrt{6}\left(1\mp x_\pm\right)^3y^2}{6\sqrt{6}\left(1\mp x_\pm\right)^3y^2+x_\pm\left[(1\mp x_\pm)^2y^2+(1-2y)x_\pm^2\right]}.\label{surf}\eea The conditions $z\leq z^\pm_\text{surf}(x_\pm,y)$ follow from the physical requirement that $\Omega_m\geq 0$.


The EoS of the cubic galileon $\omega_\phi=p_\phi/\rho_\phi$ reads

\bea \omega_\phi=\frac{x_\pm(1\mp x_\pm)^2y^2\left[z+2(1-z)\left(\frac{\ddot\phi}{H^2}\right)_\pm\right]-x^3_\pm(1-y)^2z}{(1\mp x_\pm)^2y^2\left[x_\pm z-6\sqrt{6}(1\mp x_\pm)(1-z)\right]+x^3_\pm(1-y)^2z},\label{omephi}
\eea and the deceleration parameter $q=-1-\dot H/H^2$ takes the form

\bea q=\frac{1}{2}(1+3\omega_m\Omega_m)+\frac{3}{2}\left(\frac{1\mp x_\pm}{x_\pm}\right)^2-\frac{3}{2}\left(\frac{1-y}{y}\right)^2+3\left(\frac{1\mp x_\pm}{x_\pm}\right)^2\left(\frac{1-z}{z}\right)\left(\frac{\ddot\phi}{H^2}\right)_\pm.\label{despar}\eea

It is important to mention that, in general, the system of equations \eqref{ode1}-\eqref{ode3} is not closed unless the interacting function $Q$ can be expressed in terms of the dynamical variables \eqref{n-var}. The form of such a coupling term has been discussed in the literature within the context of inflation and late-time acceleration. In particular in this work we will consider three well-known forms, namely $Q_1=3 \alpha H \rho_m$ \cite{Coley2000,Bohmer2008,Wang2005,Riess2016,Aghanim2016,Li2020,Mifsud2019,Baldi2011,Li2019},  $Q_2=3\beta\rho_m\dot{\phi}$ \cite{Coley2000,Bohmer2008, Zhang2005,Tocchini2002,Costa2015,Cicoli2012,Barros2019} and  $Q_3=3\epsilon H\rho_\phi$ \cite{Pavon2008,Yang2018,Pan2019,Yang2020,Yang2021,Kundu2021,Cardenas2019}. The corresponding analysis will be performed below in separate subsections.


\subsection{Scenario 1: $Q= 3\alpha H\rho_m$}

In this subsection we consider the interacting term 

\bea Q =3 \alpha H\rho_m \,\,\, \Rightarrow \,\,\,\frac{Q}{H^3}=9\alpha \Omega_m,\label{q1-term}\eea where $\alpha$ is a coupling constant \cite{Coley2000,Bohmer2008,Wang2005,Riess2016,Aghanim2016,Li2020,Mifsud2019,Baldi2011,Li2019}. Equation \eqref{q1-term} is to be substituted back into Eq. \eqref{ddotphi} in order to write the dynamical system \eqref{ode1}-\eqref{ode3} in closed form.

For a quintessence scalar field this scenario possesses a stable attractor solution and can thus provide a natural solution to the cosmic coincidence problem. In \cite{Wang2005} it was found that interacting dark energy can alleviate the current discrepancy between the value of the Hubble constant $H_0$ inferred from measurements of CMB anisotropies obtained from the Planck satellite and from the local measurement based on the method of distance ladder \cite{Riess2016,Aghanim2016}. The effects  of the possible excess of the 21-cm line global signal at the epoch of cosmic dawn have been worked out in \cite{Li2020} and in \cite{Mifsud2019}. The authors performed a global analysis in order to constrain an interacting dark energy model compatible with current data from Planck observations. A detailed numerical study to calculate the growth rate of density perturbations in early universe was performed in \cite{Baldi2011}. Including the JLA sample of type Ia SN observations, the Planck 2015 distance priors data of CMB anisotropies, the BAO data, and the Hubble constant direct measurement, in \cite{Li2019} the authors constrained typical interaction forms and determined a constant parameter $\alpha$ located around the interval $\{-0.155^{+.488}_{-.178}, 0.275^{+0.058}_{-0.31}\}$, depending of the set of experimental data.

\begin{figure}[ht]\centering
\includegraphics[width=8cm]{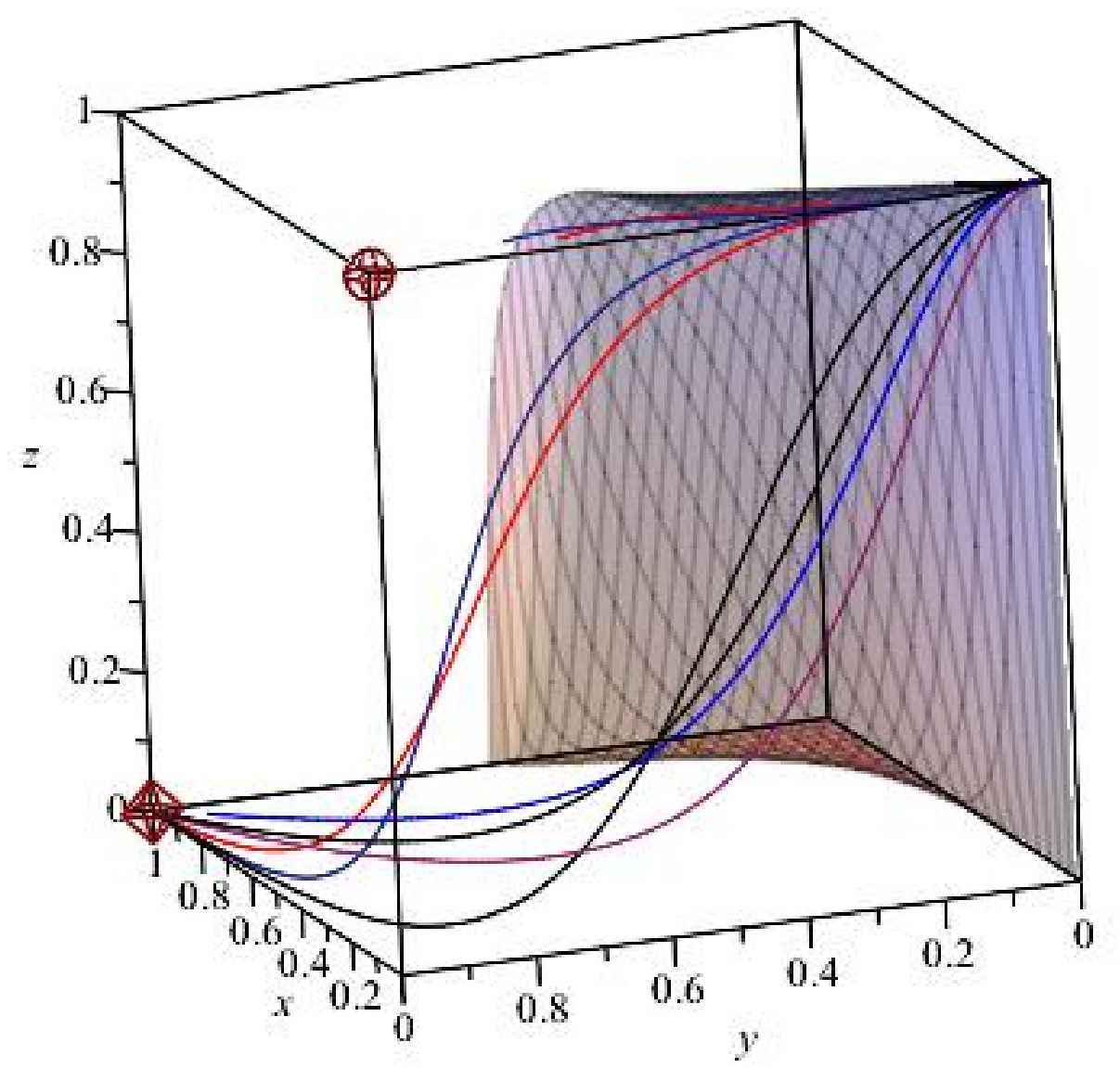}
\includegraphics[width=8cm]{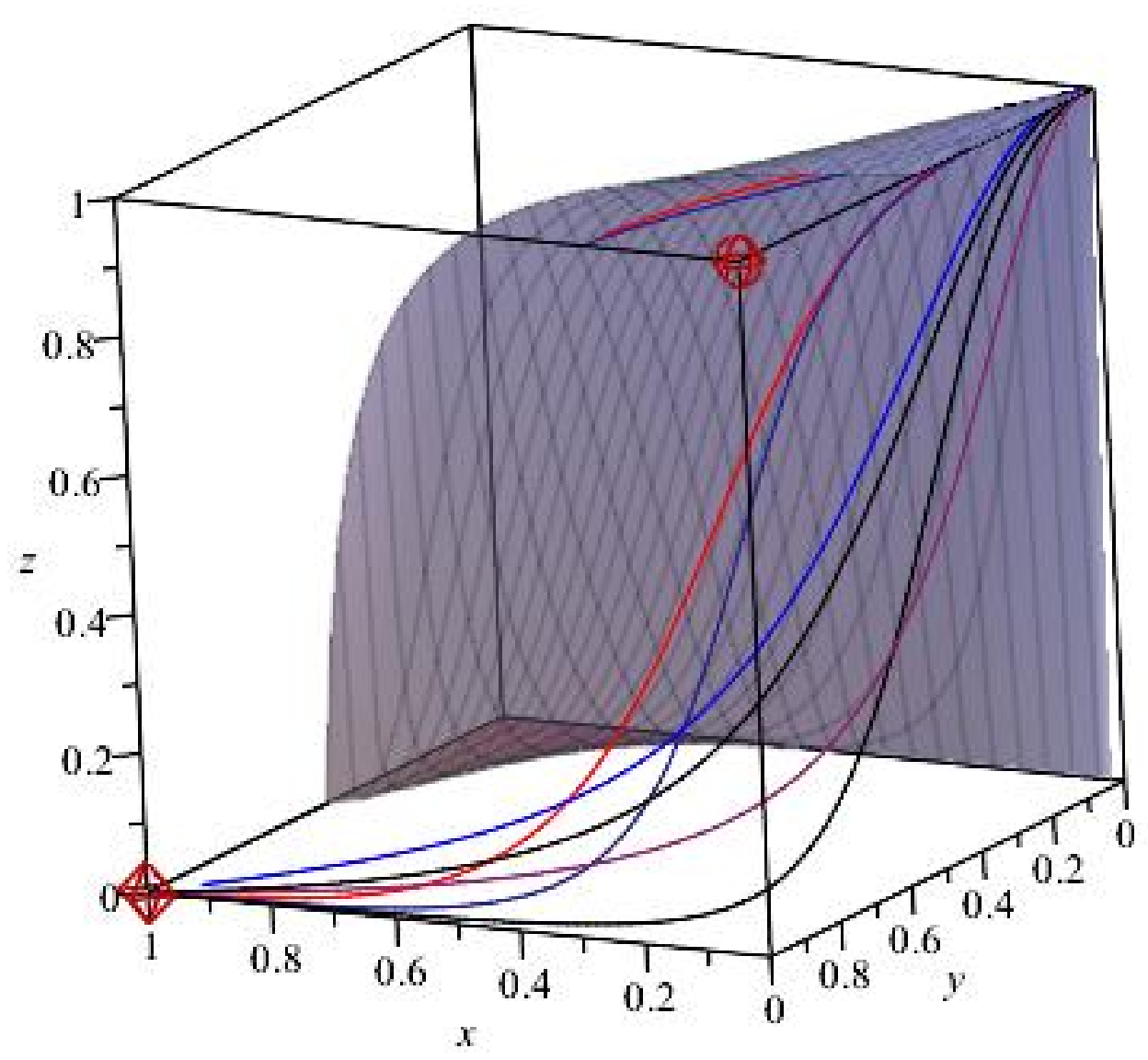}
\includegraphics[width=8cm]{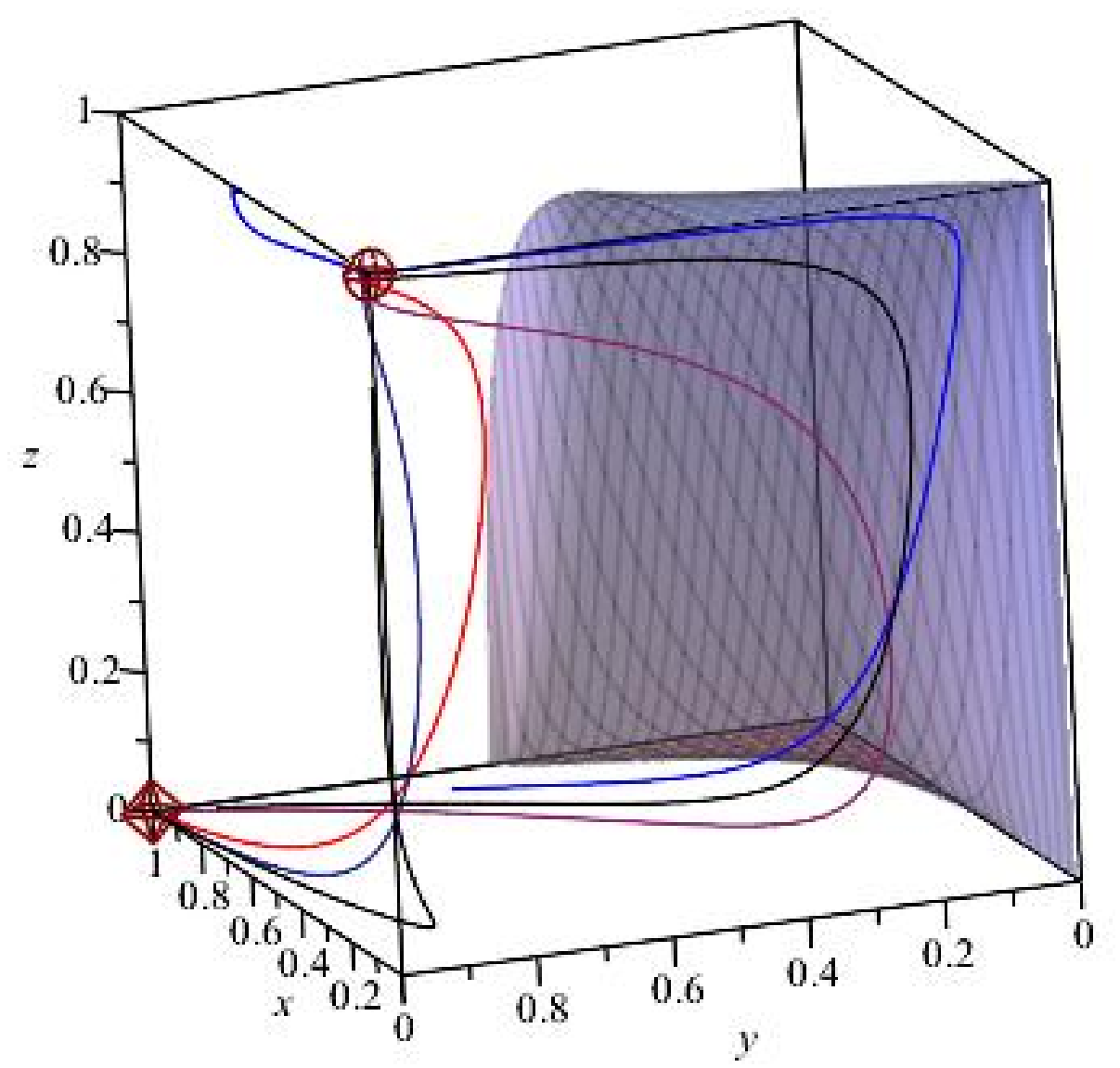}
\includegraphics[width=8cm]{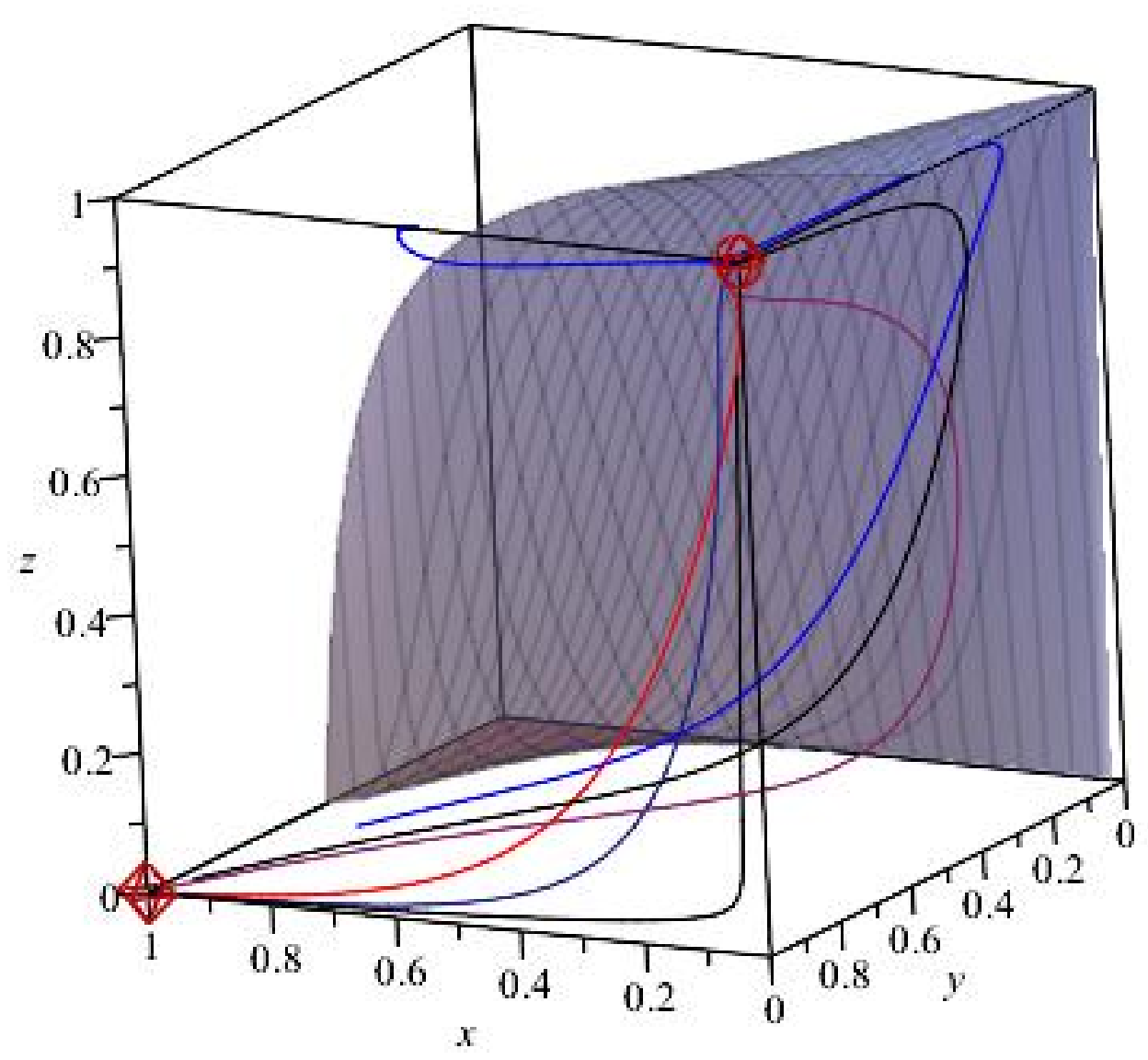}
\caption{Drawing of the 3D phase portrait for the interacting term $Q=3\alpha H\rho_m$, for two different orientations $[\theta,\varphi,\psi]$, where $theta$, $\varphi$ and $\psi$ are the Euler angles. In the left panels the orientation is $[150,75,-10]$, while in the right panels it is $[110,75,-10]$. We show drawings for two different choices of the free parameters $[\alpha,\lambda]$. In the top panels we chose $[0,0]$, while in the bottom panels we set these parameters to $[1,1]$. The gray-bluish surface Eq. \eqref{surf}, represents the boundary of the physically meaningful subspace of the phase space. It is apparent that the bigbang solution associated with point $P_1:(x=1,y=1,z=0)$, is the past global attractor. Meanwhile, the critical point $P_{11}:(x=0,y=1,z=1)$, corresponding to stiff-matter solution, is a future local attractor for a set of trajectories generated by given initial conditions while for other trajectories it is a saddle critical point instead.}\label{Figq1}
\end{figure}

The physically meaningful critical points $P^\pm:(x_\pm,y,z)$ of the dynamical system \eqref{ode1}-\eqref{ode3}, i.e., those found in the phase space $\Psi$ \eqref{psi}, together with their existence conditions are presented in Table \ref{tab-1}. In addition, for each critical point we computed the values of the EoS $\omega_{\phi}$, the DM dimensionless energy density $\Omega_m$ and the deceleration parameter $q$. We also determine the existence and stability properties of these points. The results are summarized in TAB. \ref{tab-2}. For illustration purpose we also include FIG. \ref{Figq1}, where the three-dimensional phase is drawn for given values of the free constants $\alpha$, $\lambda$.


\begin{table*}\centering
\begin{tabular}{||c||c|c|c|c|c|c|c||}
\hline\hline
Crit. Point        & $x_\pm$   &   $y$   &    $z$     & Existence  & $\Omega_m$ &  $\omega_{\phi}$   & $q$ \\
\hline\hline
$P_{1}^{\pm}$     & $ \pm 1$   &    1    &    $0$      & always &  1 & und.   &     1/2  \\
\hline
$P_2^{\pm}$     & $ \pm 1$   &    0    &    $0$      & always  &   und.  & $-1$   &     und.  \\
\hline
$P_3$     & 0   &    1    &    0      & always  &   und.  & 0   &  $2-3\alpha$  \\
\hline
$P_{4}$  & 0   &   0 &  0 & always  &  und. & 0 & $2-3\alpha$ \\
\hline
$P_{5}^{\pm}$     & $ \pm 1$   & $1/2$   &    $1$  & $\lambda=0$  & 0  & $-1$ & $-1$    \\
\hline
$P_{6}^{\pm}$     & $\pm 1/2$   & $1$   &    $1$       & always       & 0  & $1$ & $2$  \\
\hline
$P_{7}^{\pm}$     & $ \frac{ 1 \mp \sqrt{-\alpha}}{1+\alpha}$   & $1$   &    $1$       & $-1 < \alpha  \leq 0 $   &  $1+\alpha$  & 1 & $\frac{1}{2}(1-3\alpha)$  \\
\hline
$P_{8}$ & $\frac{\sqrt{6}}{\lambda + \sqrt{6}}$  & $\frac{\sqrt{6}}{\sqrt{6}+\sqrt{6-\lambda^2}}$   &   $1$ & $\lambda \leq \sqrt{6}$ & 0 & $\frac{\lambda^2}{3}-1$  &     $\frac{\lambda^2}{2}-1$ \\
\hline
$P_{9}$  & $\frac{2\lambda}{2\lambda + \sqrt{6}(1-\alpha)}$   &   $\frac{2\lambda}{2\lambda + \sqrt{6(\alpha-1)^2 + 4 \alpha \lambda^2}}$    &  1 & $ \alpha^2 \leq 1$ &  $\frac{(1-\alpha)(\lambda^2+3\alpha-3)}{\lambda^2}$  &  $\frac{-\alpha\lambda^2}{\alpha(\lambda^2-6)+3(\alpha^2+1)}$ & $\frac{1}{2}(1-3\alpha)$ \\
\hline
$P_{10}$  & 0   &  0 &  1 & always  &  und. & -1 & und. \\
\hline
$P_{11}$  & 0   &  1 &  1 & always  &  und. & 1 & und. \\
\hline\hline
\end{tabular}\caption{Scenario 1: $Q=3\alpha H\rho_m$. Physically meaningful critical points of the autonomous system \eqref{ode1}-\eqref{ode3}, their existence conditions as well as several physically meaningful parameters are summarised.}\label{tab-1}\end{table*}



\begin{table*}\centering
\begin{tabular}{||c||c|c|c|c||}
\hline\hline
Crit. Point   & $\lambda_1$   &   $\lambda_2$  &    $\lambda_3$     & Stability    \\
\hline\hline
$P_{1}^{\pm}$  &      und.  &   und.   &    $3/2$   &   unstable node  (numerical analysis)     \\
\hline
$P_{2}^{\pm}$  &      und.  &   und.   &    und.   &   saddle  (numerical analysis)     \\
\hline
$P_{3}$  &      und.  &   und.   &    $3(\alpha-1)$   &   saddle  (numerical analysis)     \\
\hline
$P_{4}$  &  $3(\alpha-1)$ & und. & und. & saddle (numerical analysis) \\
\hline
$P_{5}^{\pm}$  &   $0$  &  $-3$   &  $\frac{3}{2}(\alpha-2)$  & stable if $\alpha<2$ (numerical analysis) \\
&               &           &       &  saddle otherwise \\
\hline
$P_{6}^+$  &   $-6$   &  $3(1+\alpha)$  &  $3 - \sqrt{\frac{3}{2}}\lambda$ & stable if $\alpha<-1$ and $\lambda>\sqrt{6}$ \\
&   &   &   &  saddle if either $\alpha>-1$ or $\lambda<\sqrt{6}$ or both \\
\hline
$P_{6}^-$  &   $-6$   &  $3(1+\alpha)$  &  $3 + \sqrt{\frac{3}{2}}\lambda$ &  saddle \\
\hline
$P_{7}^+$ & $3(\alpha-1)$ & $\frac{1}{2}\left(3-3\alpha - \sqrt{-6\alpha}\lambda\right)$ & $-3(\alpha+1)$ & stable if $\lambda> \frac{1-\alpha}{\sqrt{-2 \alpha/3}}$, saddle otherwise \\
\hline
$P_{7}^-$ & $3(\alpha-1)$ & $\frac{1}{2}\left(3-3\alpha + \sqrt{-6\alpha}\lambda\right)$ & $-3(\alpha+1)$ &  saddle \\
\hline
$P_{8}$  &  $-\lambda^2$ &  $\lambda^2+3\alpha-3$ &  $\frac{\lambda^2}{2}-3$ & stable if $\alpha<1-\lambda^2/3 $, saddle otherwise \\
\hline
$P_{9}$  &  $3(\alpha-1)$ & $ A +  \sqrt{B}$ & $ A - \sqrt{B}$ & see Fig.\ref{Fig1} \\
\hline
$P_{10}$  &  0 & 0 & 0 & saddle (numerical analysis) \\
\hline
$P_{11}$  &  0 & 0 & 0 & stable or saddle depending on the\\
&&&& free parameters and on the\\
&&&&initial conditions (numerical analysis) \\
\hline\hline
\end{tabular}\caption{Scenario 1: Interacting term $Q=3\alpha H\rho_m$. The eigenvalues of the linearization matrix and the stability properties of the physically meaningful critical points are shown. Here we define $A:= \alpha\left[\frac{9\alpha-2\lambda ^2-9}{2(1-\alpha )}\right]-\frac{3}{2}$, and $B:= \frac{1}{16} \left[\frac{4 \alpha ^2 \lambda ^4}{(\alpha -1)^2}+\frac{648 (\alpha -1)^4}{\lambda ^4}+\frac{432 (\alpha -1)^3}{\lambda ^2}-\frac{12 \alpha  (\alpha +3) \lambda ^2}{\alpha -1}+9 (\alpha -26) \alpha +81\right]$.}\label{tab-2}\end{table*}


Let us briefly comment on the most salient features of the critical points.

\begin{enumerate}
\item The matter-dominated solution $P_1^\pm:(\pm 1,1,0)$ exists independent of the interacting term. In this case $z=0$ $\Leftrightarrow$ $H^2\gg\sigma^{-1}$, which means that this solution must be associated with a cosmological bigbang singularity. It is always an unstable node that marks the origin of phase space trajectories.

\item The solution $P_2^\pm:(\pm 1,0,0)$ is associated with a saddle critical point and thus it is associated with a transient solution. The energy density of the scalar field is the negative of the pressure leaving  $\omega_{\phi}=-1$, so that the galileon behaves as a cosmological constant. This point exists always and it may be associated with early time inflation since $H^2\gg\sigma^{-1}$. 

\item The  solution $P_3:(0,1,0)$ is a saddle critical point in the equivalent phase space. The pressure of the scalar field vanishes leaving  $\omega_{\phi}=0$, so that, in this case the galileon behaves as dust. This point always exists and it depicts accelerated expansion if $\alpha>2/3$. 

\item The solution $P_4:(0,0,0)$ is a saddle critical point and thus it represents a transient stage of the cosmic expansion. The equation of state parameter $\omega_{\phi}$ is undetermined. This point always exists and it is accelerated whenever $\alpha >2/3$. We have that,

\bea H^2\gg\sigma^{-1},\;\dot\phi^2\gg H^2,\;V\gg H^2,\nonumber\eea so that it is an early time solution that is dominated by the energy density of the galileon.


\item The point $P_5^\pm:(\pm 1,1/2,1)$ corresponds to late time de Sitter solution since $H_0\ll\sigma^{-1}$. It exists in the cubic galileon and in the standard interacting quintessence cosmology only if the potential vanishes. Since one of the eigenvalues of the Jacobian vanishes, the second one is negative and the remaining eigenvalue is negative for $\alpha < 2$, in order to determine the stability we must either find a Liapunov function and then to apply the centre manifold theorem. Alternatively, we must to numerically evaluate the solution. In the present case, as result of the numeric study, we found that the solution is stable if $\alpha<2$. Otherwise, it is a saddle equilibrium point. This represents a modification with respect to the cubic galileon and to the interacting quintessence, where the solution is always stable. 


\item The super-decelerated stiff-matter solution $P_6^\pm:(\pm 1/2,1,1)$ is related with a scalar field's kinetic energy density dominated universe. It exists in the cubic galileon and in the standard interacting quintessence for all values of $\lambda$ and $\alpha$. However, the phenomenological properties depend on the specific form of the potential and on the interacting term. The stability properties of this critical point are summarised in TAB. \ref{tab-2}.


\item The points $P_7^\pm: (\frac{1 \mp\sqrt{-\alpha}}{1+\alpha},1,1)$ emerge only if the interacting coupling constan $\alpha$ is in the interval $-1\leq\alpha\leq 0$. The deceleration parameter is given by $q=(1-3\alpha)/2$, so that it corresponds to decelerated scaling solution with equation of state parameter $\omega_{\phi}=1+\alpha$. The existence, stability and other relevant properties are summarised in TABs. \ref{tab-1} and \ref{tab-2}. It follows that,

\bea \frac{\dot H}{H^2}&&=-\frac{3}{2}(1-\alpha)\,\,\,\Rightarrow\,\,\,H(t)=\frac{2}{3(1-\alpha)(t-t_0)},\,\,\,\,\,\textrm{for}\,\,\,\,\,(t \geq t_0),\nonumber \\
&&\Rightarrow a=a_0(t-t_0)^{\frac{2}{3(1-\alpha)}},\nonumber\eea where $t_0$ is an integration constant. In the limit $t \rightarrow \infty$, we have $a \rightarrow \infty$.


\item The point $P_8:(\sqrt{6}/(\lambda + \sqrt{6}), \sqrt{6}/(\sqrt{6}+\sqrt{6-\lambda^2}),1)$ corresponds to the scalar field dominated solution. It represents scaling between the scalar field's kinetic energy and its potential energy density. This point can be the late time state of the universe. Its existence is related to the concrete form of the self-interacting potential by the relation $\lambda\leq\sqrt{6}$, and it is accelerated for $\lambda^2<2$. It is  a stable node for $ \alpha\leq (1-\lambda^2/(3 \lambda)$, otherwise it is a saddle.


\item The point $P_9:(2\lambda/[2\lambda+\sqrt{6}(1-\alpha)], 2\lambda/[2\lambda+\sqrt{6(\alpha-1)^2+4\alpha\lambda^2}],1)$  is closely related to the concrete form of the self-interacting potential and of the interacting term. In this case neither the scalar field nor the matter entirely dominate the universe. It arises both in the cubic galileon and the interacting quintessence scenario. The solution exists only whenever $\alpha\leq 1$ and $\lambda\neq 0$. It is accelerated if $\alpha>1/3$. The existence, stability and other relevant properties of the solution are shown in TABs. \ref{tab-1} and \ref{tab-2}. 


\item The solution $P_{10}:(0,0,1)$ is a saddle critical point in the equivalent phase space. The energy density of the scalar field equals the negative of its parametric pressure, thus leading to $\omega_{\phi}=-1$. This point always exists. However, we cannot say anything about its stability since it is a non-hyperbolic critical point. Applying numerical analysis we can determine that it is a saddle point. This is a late-time solution since $H^2\ll\sigma^{-1}$.


\item The late-time ponit $P_{11}:(0,1,1)$ posseses a EoS parameter $\omega_{\phi} =1$, so that it corresponds to a stiff matter dominated solution. As in the previous case we can say nothing about its stability since it is a non-hyperbolic critical point. However, numerically it is found that it is either a saddle or a local attractor depending on the free parameters $\alpha$, $\lambda$ and on the initial conditions.

\end{enumerate}

The above results generalise those obtained in \cite{Leon2013}, \cite{Coley2000} and \cite{Bohmer2008} by means of a slightly different procedure.


\begin{figure}\centering
\includegraphics[width=8cm]{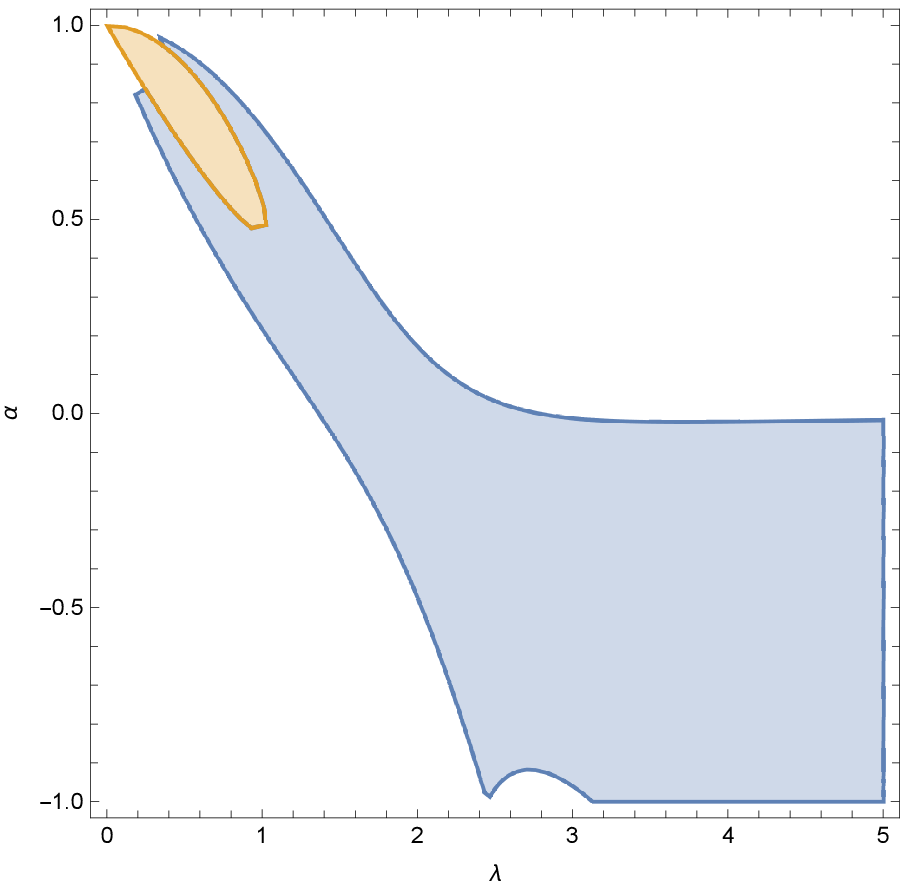}
\caption{Regions of stability for the solution $P_9$ in the $(\lambda,\alpha)$ parameter plane. Here the blue-region denotes a stable node  while the brown-shaded one is related to a stable spiral. The white surface corresponds to a saddle behaviour.}\label{Fig1}
\end{figure}


\subsection{Scenario 2: $Q=3\beta\rho_m\dot{\phi}$}

In this subsection we consider the interacting function 

\bea Q=3\beta\rho_m\dot{\phi}\;\Rightarrow\;\frac{Q}{H^3}=9\sqrt{6}\beta\Omega_m\left(\frac{1 \mp x}{x}\right),\label{q2-term}\eea where $\beta$ is the coupling parameter that characterises the strength of the non-minimal coupling  \cite{Coley2000,Bohmer2008, Zhang2005,Tocchini2002,Costa2015,Cicoli2012,Barros2019}. This scenario provides a natural solution to the cosmic coincidence problem for a stable attractor solution  consistent with standard nucleosynthesis \cite{Zhang2005,Tocchini2002}. In \cite{Costa2015} the authors put constraints on this model using the CMB measurements from the Planck satellite together with BAO, SNIa and $H_0$ data for a quintessence model for dark energy interacting with dark matter via a Yukawa interaction. A dynamical analysis is carried out in \cite{Cicoli2012} for a model of quintessence in string theory where the light rolling scalar is radiatively stable and couples to Standard Model matter. In \cite{Barros2019}  using $f \sigma_8$ data  from  redshift  space  distortions the authors found that the best fit for the coupling constant is $\beta = 0.079^{+.059}_{-0.067}$.

The physically meaningful critical points $R^\pm:(x_\pm,y,z)$ of the dynamical system \eqref{ode1}-\eqref{ode3}, which are found in the phase space $\Psi$ defined in Eq. \eqref{psi}, together with their existence conditions are presented in Table \ref{tab-3}. The values of several cosmological parameters are also shown. Finally, we determine the stability of these solutions. These results are summarised in Table \ref{tab-4}.


\begin{table*}\centering
\begin{tabular}{||c||c|c|c|c|c|c|c||}
\hline\hline
Crit. Point        & $x_\pm$   &   $y$   &    $z$     & Existence & $\Omega_m$ &  $\omega_{\phi}$   & $q$ \\
\hline\hline
$R_{1}^{\pm}$     & $\pm 1$   &    1    &    $0$      & always  & 1&     und.   &     1/2  \\
\hline
\hline
$R_{2}^{\pm}$     & $\pm 1$   &    0    &    $0$      & always  & und. &     und.   &   und.  \\
\hline
$R_{3}^{\pm}$   & $\frac{\sqrt{6}\beta}{1 \pm \sqrt{6}\beta}$  & 0 & $0$  & $ 0 \leq \beta \leq \frac{1}{2\sqrt{6}}$  & und. &  0  & -1  \\
\hline
$R_{4}^{\pm}$     & $\frac{\sqrt{6}\beta}{1 \pm \sqrt{6}\beta}$  &    1  &    $0$      & $\beta \geq 0$  & und. & 0 & -1  \\
\hline
$R_{5}^{\pm}$   & $ \pm 1$   & $1/2$   &    $1$  & always & 0  & $-1$ & $-1$ \\
\hline
$R_{6}^\pm$ & $\pm 1/2$   & $1$   &    $1$  & always  & 0  & $1$ & $2$ \\
\hline
$R_{7}$ & $ \frac{1}{1-\sqrt{6}\beta}$  & $1$ & $1$  & $0\leq\beta\leq\frac{1}{\sqrt{6}}$  & $1-6\beta^2$  & $1$ & $\frac{1}{2}+9\beta^2$ \\
\hline
$R_{8}$ & $\frac{\sqrt{6}}{\lambda+\sqrt{6}}$  & $\frac{\sqrt{6}}{\sqrt{6}+\sqrt{6-\lambda^2}}$ & $1$ & $\lambda \leq \sqrt{6}$ &  $0$ &  $\frac{\lambda^2}{3}-1$ &     $\frac{\lambda^2}{2}-1$\\
\hline
$R_{9}$ & $\frac{2(\lambda+3\beta)}{\sqrt{6}+2(\lambda+3\beta)}$ & $\frac{\lambda+3\beta}{\lambda+3\beta+\sqrt{3/2+3\beta(\lambda+3\beta)}}$ & $1$ & $\lambda \leq \sqrt{6} \,\, \& \, \, \beta \geq \frac{3-\lambda ^2}{3 \lambda } \,\,\, \textrm{or}$ & $\frac{\lambda^2+3\beta \lambda-3}{(\lambda+3\beta)^2}$ & $\frac{-\beta(\lambda+3\beta)}{3\beta^2+\lambda \beta +1}$  &  $\frac{\lambda -6\beta}{2(\lambda+3\beta)}$ \\
&&&&$ \lambda >\sqrt{6}\,\, \& \,\, \beta \geq \frac{\sqrt{\lambda ^2-6}}{6}-\frac{\lambda }{6}$ & & & \\
\hline
$R_{10}$     & 0   &  0   &   1  & always   & und.  & -1 &     und. \\ 
\hline
$R_{11}$     & 0   &  1   &   1  & always   & und.  & 1 &     und.   \\
\hline\hline
\end{tabular}\caption{Scenario 2: Interacting term $Q=3\beta\rho_m\dot{\phi}$. The physically meaningful critical points of the autonomous system \eqref{ode1}-\eqref{ode3} together with their existence conditions are shown. Additionally, we write the values of several relevant cosmological parameters matter such as the density parameter $\Omega_m$, the EoS $\omega_{\phi}$ and the deceleration parameter $q$.}\label{tab-3}\end{table*}


\begin{table*}\centering
\begin{tabular}{||c||c|c|c|c||}
\hline\hline
Crit. Point   & $\lambda_1$   &   $\lambda_2$  &    $\lambda_3$     & Stability    \\
\hline\hline
$R_{1}^\pm$  &      2/3  &   $3/2$   &    und.   &   unstable node (numerical analysis)      \\
\hline
$R_{2}$  &      und.  &   und.  &   und.    &   saddle   (numerical analysis)   \\
\hline
$R_{3}$  &      und.  &   $3$   &    $3/2$   &   saddle   (numerical analysis)    \\
\hline
$R_{4}^\pm$  &  0  &   $-3$   &    $-3/2$   &    stable node (numerical analysis)       \\
\hline
$R_{5}^\pm$  &     0    &    -3    &  $-3$      & stable (numerical analysis)       \\
\hline
$R_{6}^+$  &     $-6$   &    $3 + 3 \sqrt{6} \beta$    &    $3 - \sqrt{\frac{3}{2}} \lambda $       & stable  if $\lambda>\sqrt{6}$ and $ \beta< -1/\sqrt{6}$, \\
  &&&& saddle either $\lambda < \sqrt{6}$ or $\beta > -1/\sqrt{6}$ or both \\
\hline
$R_{6}^-$  &     $-6$   &    $3 - 3 \sqrt{6} \beta$    &    $3 + \sqrt{\frac{3}{2}} \lambda $       & saddle \\
\hline
$R_{7}$  &     $9 \beta^2 - \frac{3}{2}$   &    $-3(1+6\beta^2)$    &   $\frac{3}{2}+3\beta(\lambda+3\beta)$       &   stable if $\lambda > - \frac{1+6\beta^2}{2 \beta}\,\,$, saddle otherwise  \\
\hline
$R_{8}$  &      $-\lambda^2$   &  $\frac{\lambda^2}{2}-3$  &    $\lambda^2+3\beta \lambda -3$   &  stable if  $\beta < \frac{3-\lambda^2}{3 \lambda}$, saddle otherwise \\
\hline
$R_{9}$  &  $ \frac{-3 \lambda}{\lambda+3\beta}$ &    $ C + \sqrt{D}$ & $ C - \sqrt{D} $     &  see Fig.\ref{Fig2}\\
\hline
$R_{10}$  &  0  &  0  &  0   & saddle  (numerical analysis)   \\
\hline
$R_{11}$  & 0 & 0 & 0 & stable (numerical analysis)   \\
\hline\hline
\end{tabular}\caption{Scenario 2: Interacting term $Q_2=3\beta\rho_m\dot{\phi}$. The eigenvalues of the linearization matrices and the resulting stability properties of the physically meaningful critical points are shown. Here we define $C:=\frac{27\left(3\beta\lambda+\lambda^2-3\right)[2\beta(3\beta+\lambda)+1]}{(3\beta+\lambda)^4}$ and $D:=\frac{-3\{(3\beta +\lambda)[108\beta^3(3\beta+4\lambda^2-3)+144\beta^2\lambda(\lambda^2-2)+\beta (16\lambda^4-33\lambda^2+72)+9\lambda^3-48\lambda]+72\}}{4(3\beta+\lambda)^4}$.}\label{tab-4}\end{table*}



\begin{figure}[ht]\centering
\includegraphics[width=8cm]{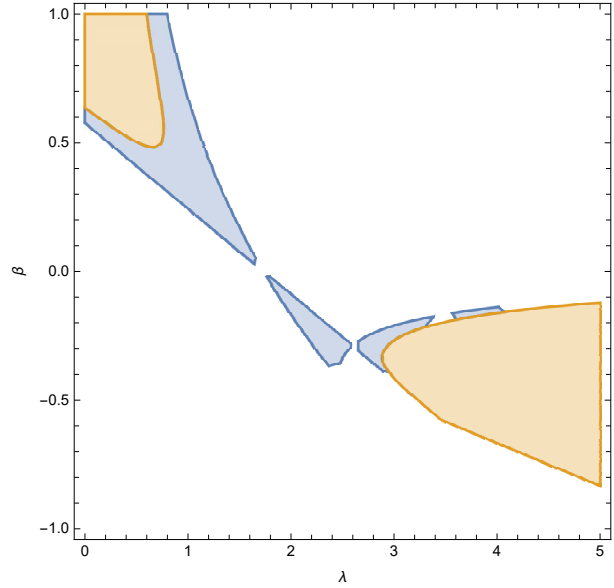}
\caption{Regions of stability for the solution $R_9$ in the $(\lambda,\beta)$ parameter plane. Here the blue-region is for the case when $R_9$ is a stable node, while the brown-shaded one is related to a stable spiral. The white surface corresponds to saddle behaviour.}\label{Fig2}
\end{figure}


\begin{figure}[ht]\centering
\includegraphics[width=8cm]{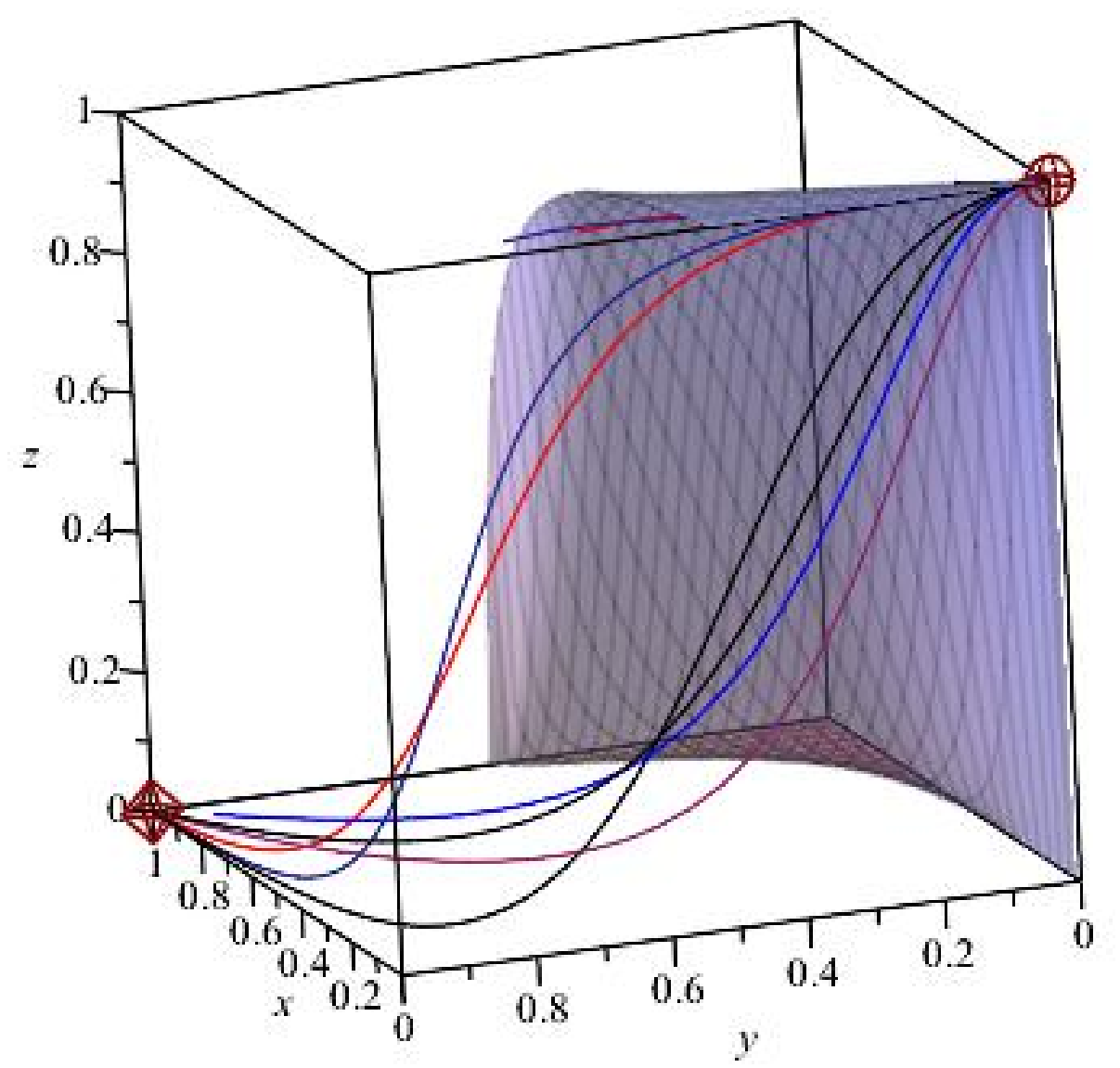}
\includegraphics[width=8cm]{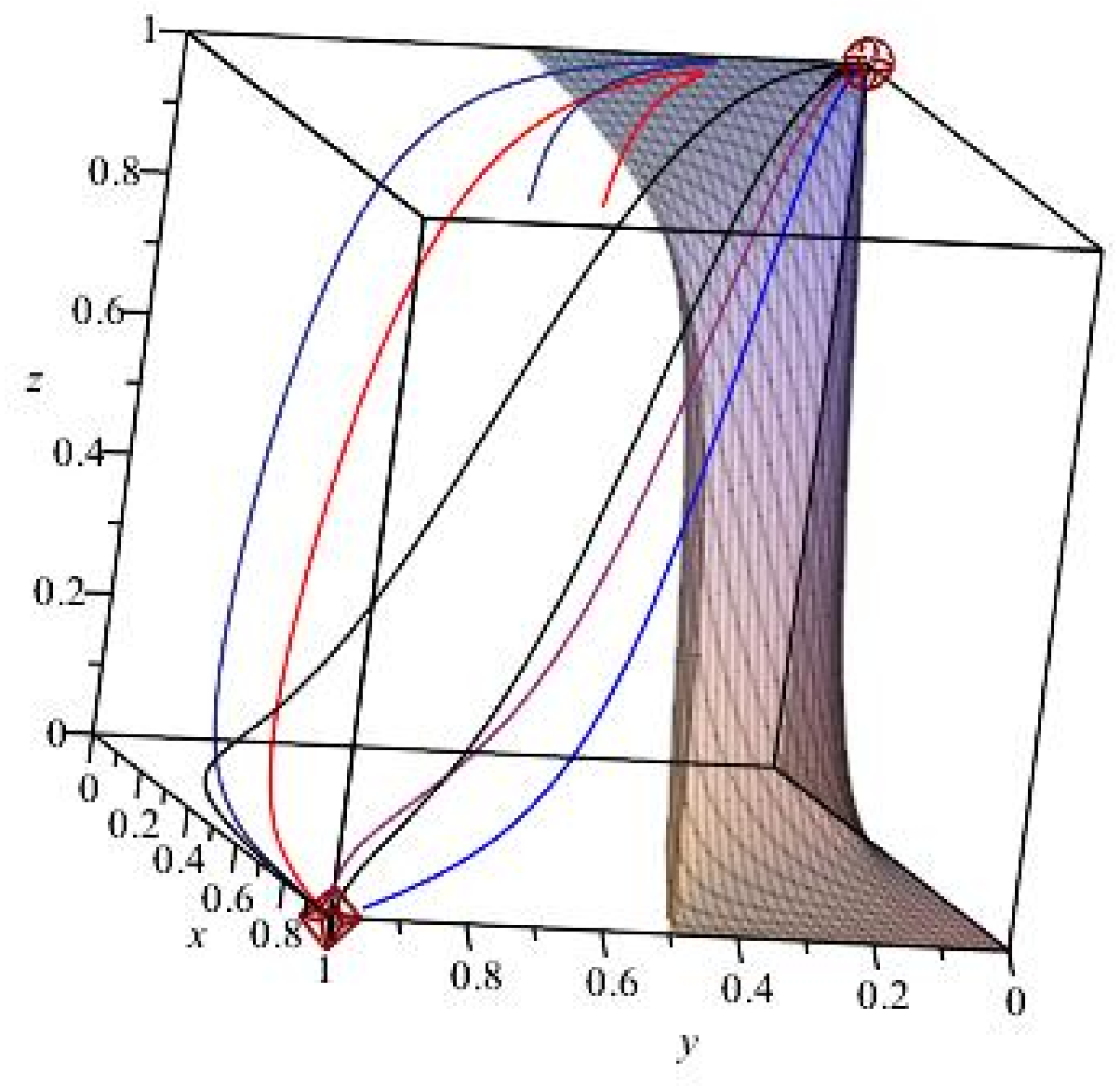}
\includegraphics[width=8cm]{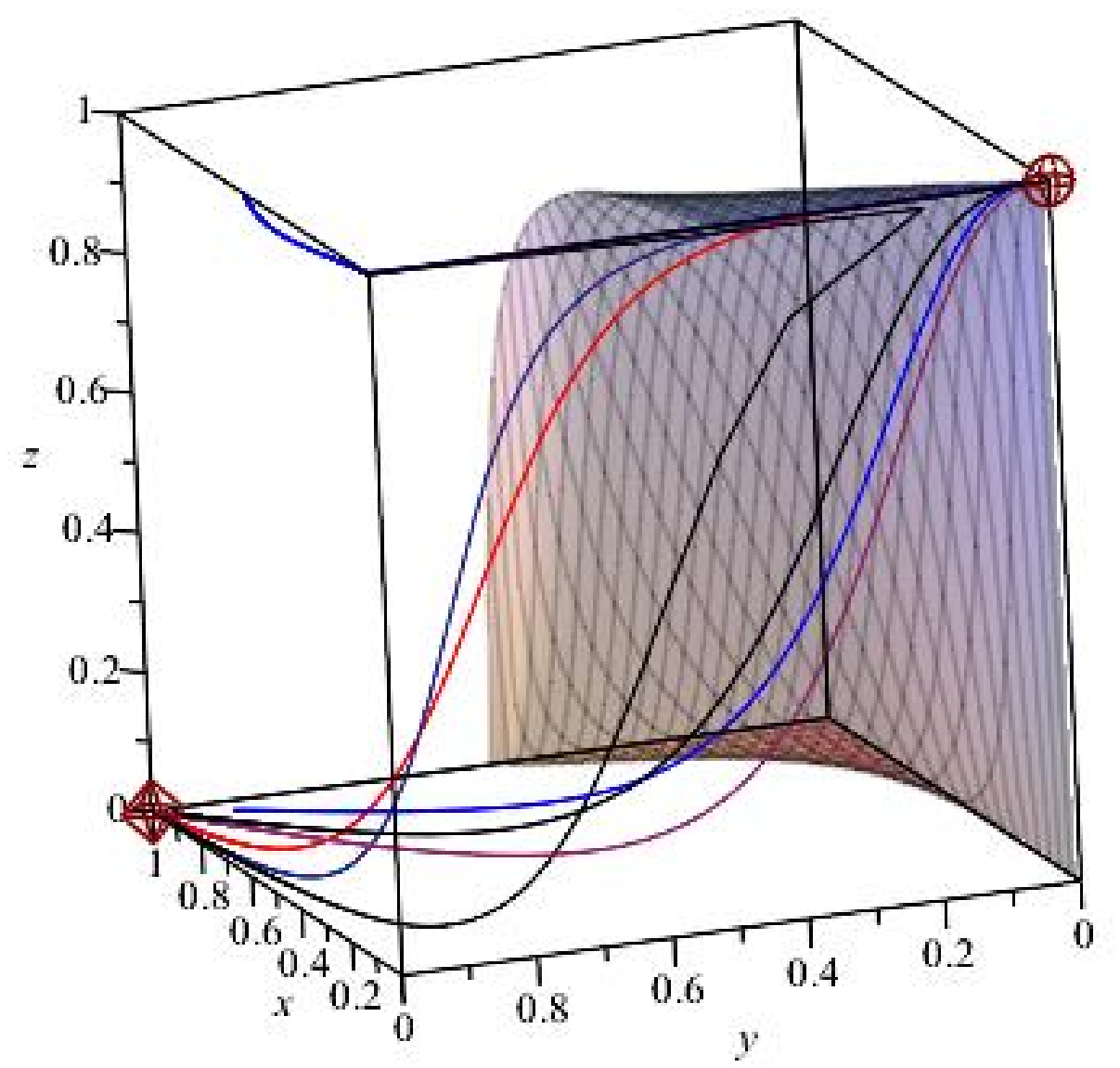}
\includegraphics[width=8cm]{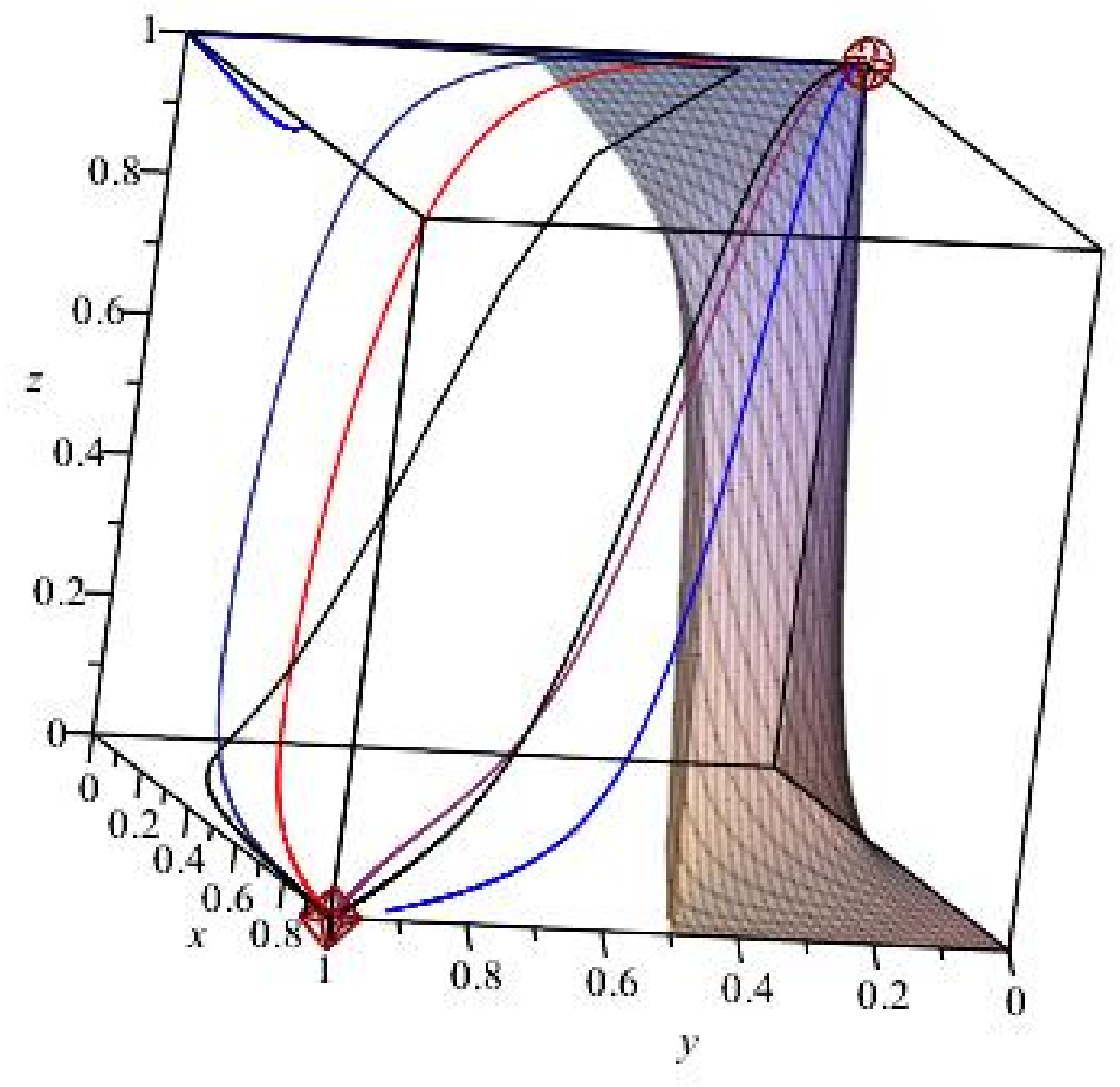}
\caption{Drawing of the 3D phase portrait for the interacting term $Q= 3\beta\rho_m\dot{\phi}$, and for two different orientations. In the left panel the orientation is $[\theta,\varphi,\psi]=[150,75,-10]$, while in the right panel $[\theta,\varphi,\psi]=[110,75,-10]$. Here we consider the same sets of values of the free parameters than in FIG. \ref{Figq1}.}\label{Figq2}\end{figure}


Below we briefly comment on the most salient features of the physically meaningful critical points.

\begin{enumerate}

\item The bigbang solution $R_1^\pm:(\pm 1,1,0)$ corresponds to the decelerated dark matter-dominated solution which is independent of the interacting coupling term. It is an unstable node (past attractor), so that it is the origin of space phase trajectories. Here $z=0$, i. e. $H^2\gg\sigma^{-1}\rightarrow\infty$, which means that this critical point may be associated with the cosmological bigbang singularity. 


\item The solution $R_2^\pm:(\pm 1,0,0)$ is a saddle critical point in the equivalent phase space, so that it represents an early transient stage of the cosmic expansion. The energy density of the scalar field vanishes thus leaving the EoS parameter undetermined. This point exists always and the deceleration parameter $q$ is undefined.


\item The dust-like solution $R_3^\pm:(\frac{\sqrt{6}\beta}{1 \pm \sqrt{6}\beta} ,0,0)$ is associated with a saddle critical point. It represents a transient epoch in the history of the universe. The pressure of the scalar field vanishes and the equation of the state parameter is zero, so that the galileon fluid behaves as dust. This solution always exists and it represents accelerated expansion. In this case,

\bea x_s=\frac{1}{\sqrt{6}\beta}\;\Rightarrow\;\dot{\phi}=\frac{H}{\beta},\;y_s\rightarrow\infty\;\Rightarrow\;V\gg H^2,\;z=0\;\Rightarrow\;H^2\gg\sigma^{-1}.\nonumber\eea 


\item The solution $R_4^\pm:(\frac{\sqrt{6}\beta}{1 \pm \sqrt{6}\beta} ,1,0)$, also represents accelerated expansion. It exists only for non-negative coupling parameter $\beta\geq 0$. This critical point is a stable node (future attractor), so that it can be the late-time state of the universe. The pressure of the scalar field vanishes so that the EoS vanishes as well. Hence, in this solution the galileon behaves as dust matter. We have that,

\bea \dot{\phi}=\frac{H}{\beta},\;H^2\gg V,\;H^2\gg\sigma^{-1}.\nonumber\eea 


\item The de Sitter point $R_5^\pm:(\pm 1,1/2,1)$, always exists. The numerical study reveals that this point is a stable node. This represents a modification with respect to the previous scenario.


\item The stiff-matter solution $R_6^\pm:(\pm1/2,1,1)$ is related with a scalar field's kinetic energy density dominated universe and exists for all values of $\beta$. It is represented by a stiff-fluid where,

\bea 3H^2=\frac{\dot\phi^2}{2},\;V\ll H^2,\;H^2\ll\sigma^{-1}.\nonumber\eea


\item The scaling solution $R_7:(1/(1-\sqrt{6}\beta ),1,1)$ represents decelerated expansion. The galileon EoS $\omega=1$. This point exists for $0\leq\beta\leq\frac{1}{\sqrt{6}}$. It is a stable node if $\lambda>-\frac{(1+6\beta^2)}{2\beta}$ and a saddle otherwise. The deceleration parameter $q=\frac{1}{2}+9\beta^2$. We have that,

\bea &&3H^2=\rho_m+\frac{\dot\phi^2}{2},\nonumber\\
&&\frac{\dot H}{H^2}=-\frac{3}{2}(1+6\beta)\;\Rightarrow\;H(t)=\frac{2}{3(1+6\beta^2)(t-t_0)},\nonumber\eea where $t_0$ is an integration constant. The last line follows from the value of the deceleration parameter in TAB. \ref{tab-3}: $q=1/2+9\beta^2$. 


\item The point $R_8:(\sqrt{6}/(\lambda+\sqrt{6}),\sqrt{6}/(\sqrt{6}+\sqrt{6-\lambda^2}),1)$ corresponds to the scalar field dominated solution, representing a scaling between the kinetic and potential energies of the scalar field. Its existence is related to the concrete form of the parameter $\lambda$. It is an accelerated solution for $\lambda^2<2$. This solution corresponds to a stable node for $\beta<(3-\lambda^2)/3\lambda$. Otherwise is a saddle.


\item The solution $R_{9}:(2(\lambda+3\beta)/[\sqrt{6}+2(\lambda+3\beta)], (\lambda+3\beta)/[\lambda+3\beta+\sqrt{3/2+3\beta(\lambda+3\beta)}],1)$ is closely related to the concrete form of the self-interacting potential and the interacting coupling term. In this case neither the scalar field nor the matter entirely dominates the universe, i.e., it is a matter-scaling solution representing scaling between the energy density of the galileon and the matter energy density. The existence and stability properties, among others, are shown in TABs. \ref{tab-3} and \ref{tab-4}.


\item A numerical investigation shows that the solution $R_{10}:(0,0,1)$ is a saddle critical point. The EoS parameter  $\omega_{\phi}=-1$, so that the galileon behaves as a vacuum fluid. In this case we have that,

\bea \dot\phi^2\gg H^2,\;V\gg H^2,\;\sigma H^2\ll 1,\nonumber\eea so that this solutions represents late-time transient stage of the cosmic expansion.


\item The solution $R_{11}:(0,1,1)$ is associated with a saddle critical point and thus it represents a transient epoch in the history of the universe. The EoS $\omega_{\phi} =1$ corresponds to a stiff matter component. As in the previous case, we can say nothing about its stability since it is a non-hyperbolic critical point. However, the numerical investigation reveals that it is a stable node (future attractor).

\end{enumerate} 

The critical points $R_7$ and $R_9$ both represent scaling solutions, however, only $R_7$ can be a late-time attractor which may alleviate the cosmic coincidence problem. With respect to the former non-minimal coupling scenario, the de Sitter solution $R_6$ is a stable node so it may represent the late-time stage of the cosmic expansion.


\begin{table*}\centering
\begin{tabular}{||c||c|c|c|c|c|c|c||}
\hline\hline
Crit. Point        & $x_\pm$   &   $y$   &    $z$     & Existence & $\Omega_m$ &  $\omega_{\phi}$   & $q$ \\
\hline\hline
$S_{1}^{\pm}$     & $\pm 1$   &    1    &    $0$      & always  & 1&     und.   &     1/2  \\
\hline
$S_{2}^\pm$     & $\pm 1$   &    0    &    $0$      & always  & und.  &     -1   &    und.  \\
\hline
$S_{3}$     & 0   &    1    &    $0$      & always  & und. &     0   &    $2 + 3 \epsilon$  \\
\hline
$S_{4}$     & 0   &    0    &    $0$      & always  & und. &     0   &    $2 + 3 \epsilon$  \\
\hline

$S_{5}^{\pm}$     & $ \pm 1$   &    1    &    $1$      & always & 1 &    1   &     1/2  \\
\hline
$S_{6}^\pm$     & $\frac{-1 \mp \sqrt{1+\epsilon}}{\epsilon}$    & $1$   &    $1$  & $-1<\epsilon<0$ & $-\epsilon$  & 1 & $2+3\epsilon/2$ \\
\hline
$S_{7}^\pm$     & $ \frac{1}{2}$    &   1 &  1      & always  &  0  &  1  & 2 \\
\hline
$S_{8}$     & 0   &    0    &  1      & always  & und.  &     -1   &    und.  \\
\hline
$S_{9}$     & 0   &    1    &   1      & always  & und.  &     1   &    und.  \\
\hline\hline
\end{tabular}\caption{Scenario 3: Interacting term $Q=3\epsilon H\rho_\phi$. The physically meaningful critical points of the autonomous system \eqref{ode1}-\eqref{ode3}, together with their existence conditions are shown. Additionally, we write the values of the normalised matter density parameter $\Omega_m$, of EoS parameter $\omega_{\phi}$ and of the deceleration parameter $q$. Here $R\equiv\sqrt{6\lambda^2-12\sqrt{6}\lambda-9\epsilon^2+36}$}\label{tab-5}\end{table*}


\begin{table*}\centering
\begin{tabular}{||c||c|c|c|c||}
\hline\hline
Crit. Point   & $\lambda_1$   &   $\lambda_2$  &    $\lambda_3$     & Stability    \\
\hline\hline
$S_{1}^\pm$  &      und.   &    und.   &   und. &  unstable node (numerical analysis)     \\
\hline
$S_{2}^\pm$  &      und.   &    und.   &   und. &  saddle (numerical analysis)     \\
\hline
$S_{3}$  &      und.   &    $6(1+\epsilon)$   &   $-3(1+\epsilon)$ &  saddle     \\
\hline
$S_{4}^\pm$  &     und.   &    $6(1+\epsilon)$  & und.      &   saddle \\
\hline
$S_{5}^\pm$  &     $-3$   &    $\frac{3}{2}$  & und.      &   saddle \\
\hline
$S_{6}^\pm$  &     $3(1+\epsilon)$   &    $-3(2+\epsilon)$    &   $\frac{1}{2}(6+3\epsilon-\sqrt{6+6 \lambda \epsilon})$       &  saddle  \\
\hline
$S_{7}^\pm$  &      -6  & 3   & 3 &  saddle     \\
\hline
$S_{8}^\pm$  &      0  & 0   & 0 &  saddle (numerical analysis)     \\
\hline
$S_{9}^\pm$  & 0   & 0   &  0 &  stable (numerical analysis)     \\
\hline\hline
\end{tabular}\caption{The eigenvalues of the linearization matrices evaluated at the critical points in TAB. \ref{tab-5} and their stability properties are shown.}\label{tab-6}\end{table*}

\begin{figure}[ht]\centering
\includegraphics[width=8cm]{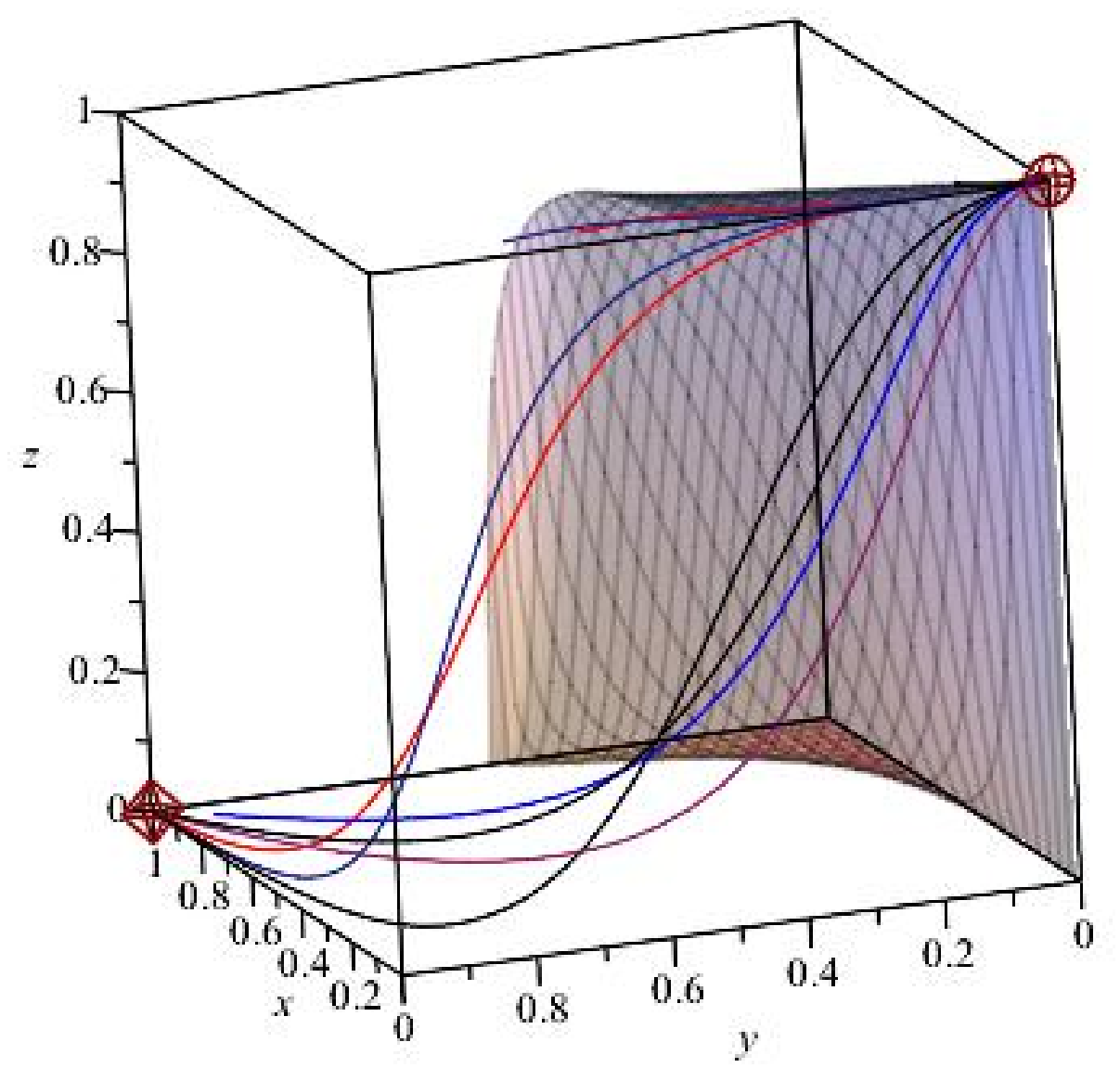}
\includegraphics[width=8cm]{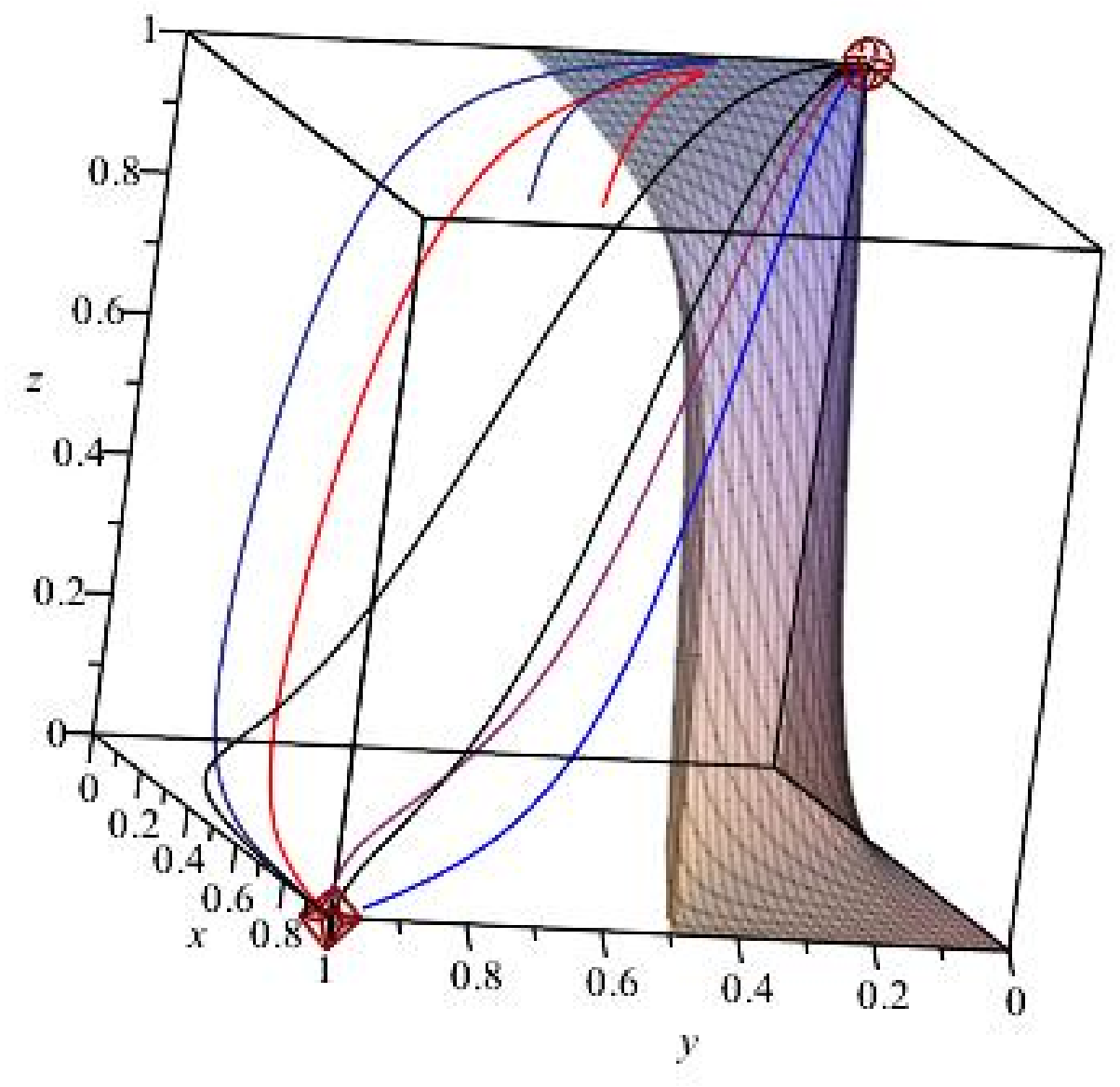}
\includegraphics[width=8cm]{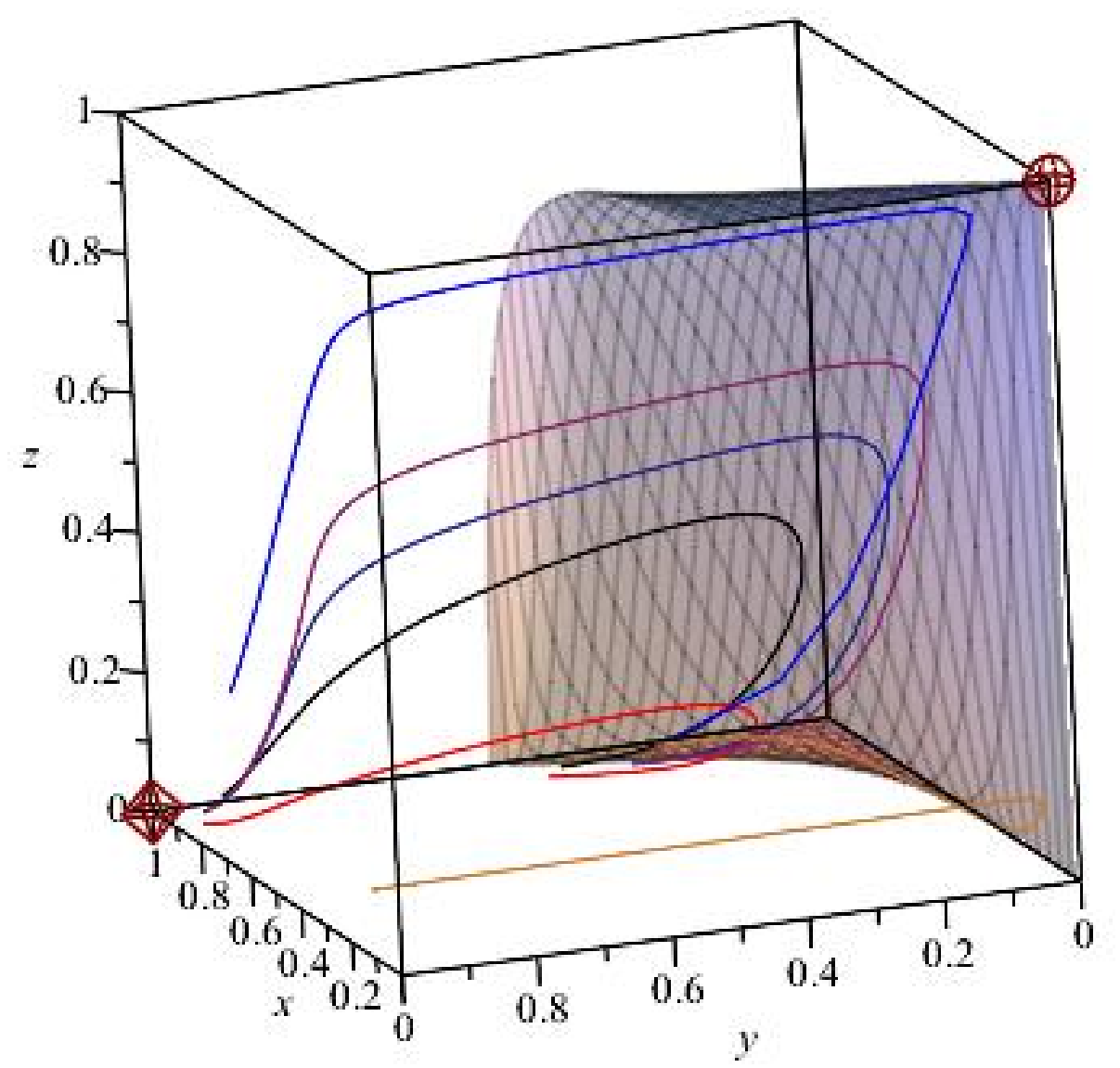}
\includegraphics[width=8cm]{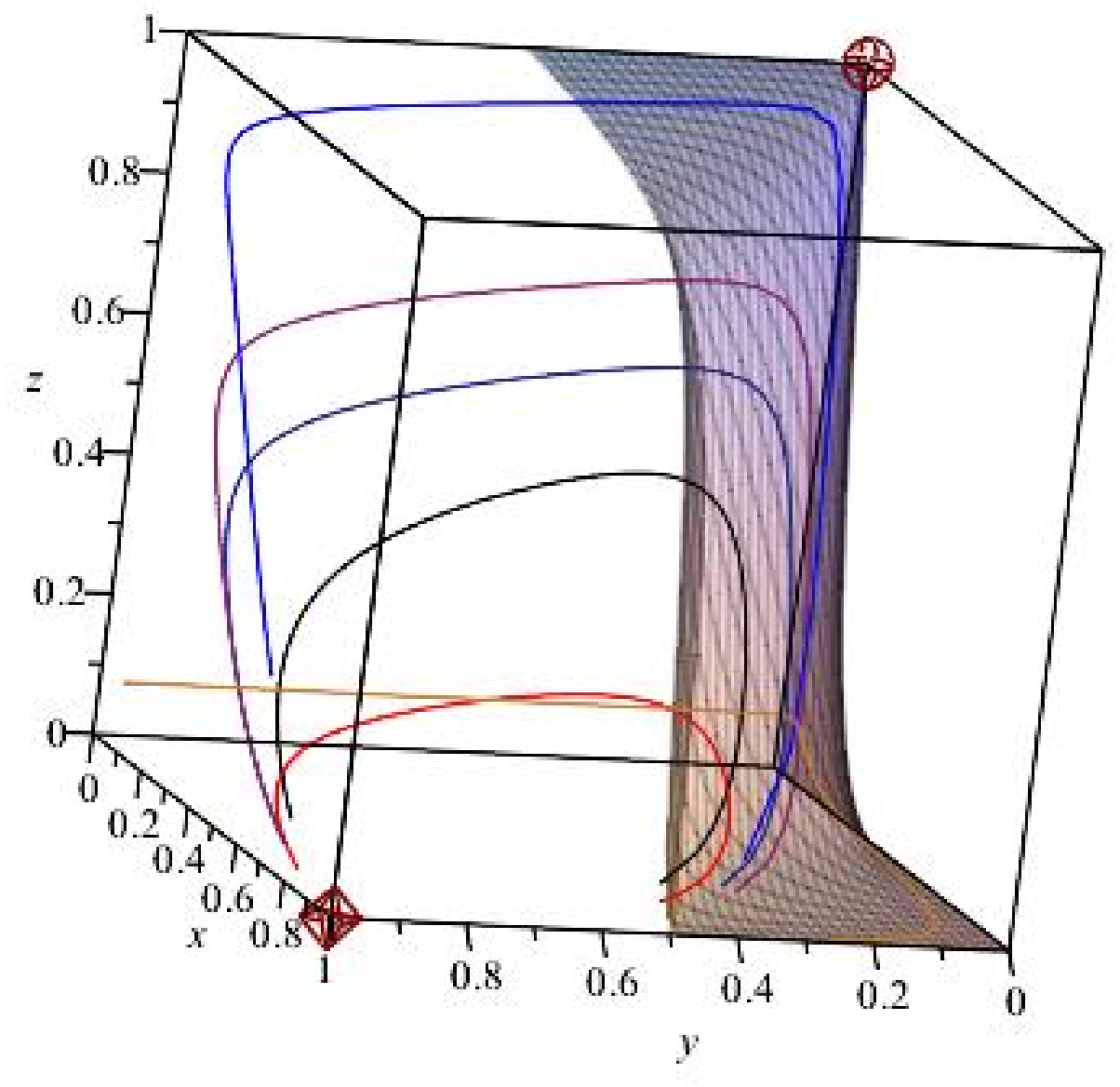}
\caption{Drawing of the 3D phase portrait for the interacting term $Q_3=3\epsilon H\rho_\phi$ for two different orientations. In the left panel the orientation is $[\theta,\varphi,\psi]=[150,75,-10]$, while in the right panel $[\theta,\varphi,\psi]=[110,75,-10]$. It is apparent that the point $P_1:(1,1,0)$ is the past global attractor, while the future (global) attractor is the critical point $P_{11}:(x=0,y=1,z=1)$.}\label{Figq3}\end{figure}

\subsection{Scenario 3: $Q_3=3\epsilon H\rho_\phi$}

In this subsection we consider the interacting function,

\bea Q=3\epsilon H\rho_\phi\;\Rightarrow\;\frac{Q}{H^3}=9\epsilon\Omega_\phi=9\epsilon(1-\Omega_m),\label{q3-term}\eea where $\epsilon$ is the coupling parameter that characterises the strength of the non-gravitational coupling to the DM \cite{Pavon2008,Yang2018,Pan2019,Yang2020,Yang2021,Kundu2021,Cardenas2019}. This model has been examined using observational data for its effect in alleviating the coincidence problem, as well as some well-known cosmological data strains \cite{Yang2018,Pan2019,Yang2020,Yang2021}. Besides, in  \cite{Kundu2021} the authors obtained the age of the universe for a spatially homogeneous interacting dark energy model for a tachyonic scalar field. They found that with the increase in the value of the coupling constant, the age of the universe is also increasing, however, beyond the limit $\epsilon=1/3,$ the age of the universe turns out to be imaginary which is a non-physical situation. Alternatively, taking into account the restrictions imposed by thermodynamics, in \cite{Cardenas2019} a study was carried out for this interaction between the dark sector and the temperature evolution for both components was reconstructed using observational data from SNIa + $H(z)$ + BAO + $f_{gas}$ + CMB. Best fit values of the cosmological parameters for the interaction provides $Q=0.07\pm 0.005$.

The physically meaningful critical points $S^\pm:(x_\pm,y,z)$ of the dynamical system \eqref{ode1}-\eqref{ode3}, which arise in the phase space $\Psi$ defined in Eq. \eqref{psi}, together with their existence conditions are presented in Table \ref{tab-5}. In addition, for each critical point we computed the values of the EoS parameter $\omega_{\phi}$, of the DM dimensionless energy density $\Omega_m$ and of the deceleration parameter $q$. We also determined the stability of these points. The results are summarised in TAB. \ref{tab-6}.

Let us briefly discuss on the physically relevant equilibrium points.

\begin{enumerate}

\item The point $S_1^\pm:(\pm 1,1,0)$ corresponds to the decelerated dark matter-dominated bigbang solution. It is always an unstable node which can be the origin of the phase space trajectories. 


\item The solution $S_2^\pm:(\pm 1,0,0)$ is a saddle critical point. The energy density of the scalar field vanishes leading to the galileon EoS to be undetermined. This point always exists.


\item The solution $S_3:(0,1,0)$ is a saddle equilibrium point that can be associated with a transient stage of the cosmic expansion. In this scenario the pressure of the scalar field vanishes so that the galileon EoS $\omega_{\phi}=0$. This point always exists and it depicts accelerated expansion if the non-minimal coupling parameter $\epsilon<-2/3$. We have that,

\bea \dot\phi^2\gg H^2,\;V\ll H^2,\;H^2\gg\sigma^{-1}.\nonumber\eea


\item The solution $S_4:(0,0,0)$ is a saddle critical point in the equivalent phase space, so that it also represents a transient stage of the expansion of the universe. The pressure of the scalar field vanishes, so that the galileon EoS $\omega_{\phi}=0$. This point always exists and it is accelerated for $\epsilon<-2/3$.


\item The decelerated DM dominated solution $S_5^\pm:(\pm 1,1,1)$ is a saddle critical point. The energy density of the scalar field matches its parametric pressure so that the EoS $\omega_\phi=1$. Hence the dark matter behaves as a stiff-fluid. This point always exists and it represents a modification with respect to the previous cases, where it only emerges for a vanishing coupling parameter.


\item The points $S_6^\pm:(\frac{-1 \mp\sqrt{1+\epsilon}}{\epsilon},1,1)$ are closely related to the concrete form of the self-interacting potential and of the interacting coupling term. In this case neither the scalar field nor the matter entirely dominate the universe. These are matter-scaling solutions which exist whenever $-1<\epsilon< 0$ and represent accelerated expansion if $\epsilon<-4/3$.


\item The stiff-matter solution $S_7:(\frac{1}{2},1,1)$ happens to be a saddle point in the phase space. This solution exists if $\epsilon=0$. We have that,

\bea3H^2=\frac{\dot\phi^2}{2},\;V\ll H^2,\;H^2\ll\sigma^{-1}.\nonumber\eea


\item The numerical investigation shows that the solution $S_8:(0,0,1)$ is a saddle critical point. The energy density of the scalar field equals the negative of its parametric pressure, so that the galileon EoS $\omega_{\phi}=-1$. This point always exists, however, we cannot say anything about its stability since it is a non-hyperbolic critical point. This is why we needed an additional numerical analysis.


\item The solution $S_9:(0,1,1)$ is a saddle critical point and thus it represents a transient epoch in the history of the universe. The EoS $\omega_{\phi}=1$, so that this equilibrium point corresponds to a stiff-matter solution. As in the previous case, we can say nothing about its stability since it is a non-hyperbolic critical point and hence, we need of an independent numerical investigation. 


\end{enumerate}

As in the previous scenarios, those critical points with $z=0$ are associated with very large values of the Hubble parameter so that these represent early times stages of the cosmic expansion. The critical points with $z=1$ must be associated with late-time dynamics instead. It can be noticed from TABs and from the illustrative FIGs. that the cubic interaction of the galileon modifies only the early time dynamics, while the non-minimal interaction modifies both, the early time and late time behaviours.


\section{Discussion}\label{sec-discuss}

In order to discuss about the impact of the non-minimal interaction of the cubic galileon and the DM, let us first summarise previous results regarding the asymptotic dynamics of the cubic galileon without the additional non-gravitational interaction with the dark matter. This will be followed by a summary of results of the interacting quintessence scenario. For simplicity and considering that the present discussion is just illustrative, we shall consider the coupling function $Q=\alpha H\rho_m$ exclusively, under the assumption of an exponential potential $V(\phi)=V_0 e^{-\lambda\phi}$. 


The critical points of the dynamical system for the cubic galileon without any additional non-gravitational interaction with the matter component, as well as their existence and stability properties are shown in TABs. 1 and 2 of reference \cite{Dearcia2016}. For self-interacting potentials beyond the exponential potential these results are summarised in \cite{Dearcia2018}. According to TAB. 1 of \cite{Dearcia2016} the physically meaningful critical points of the cubic galileon model are the following (here we use the same notation of Ref. \cite{Dearcia2016}):

\begin{enumerate}

    \item The bigbang (matter dominated) solution $P_1^\pm$, which is the global past attractor for every phase space orbit.
    
    \item The matter dominated solution $P_2^\pm$, which is a saddle equilibrium point. This means that the matter dominated phase is a transient stage in the expansion history.
    
    \item The stiff-matter solution $P_3^\pm$, which is correlated with galileon kinetic energy density dominated universe. In the present case these it is always a saddle point, while in the standard quintessence case it is the past attractor.
    
    \item The galileon dominated solution $P_4^\pm$, representing scaling between the galileon’s kinetic and potential energy densities. This can be either a saddle or a late-time attractor as in the quintessence model.
    
    \item The galileon-matter scaling solution $P_5^\pm$. Whenever it exists it is a late-time attractor. It is either a focus or a spiral equilibrium point.
    
\end{enumerate}

The critical points 2-5 where found first in Ref. \cite{Leon2013}. It is evident from the study in Refs. \cite{Leon2013,Dearcia2016,Dearcia2018} that the cubic galileon modifies only the dynamics at early times. The late-time dynamics are essentially the same as those corresponding to the quintessence model.


\begin{figure}[ht]\centering
\includegraphics[width=5cm]{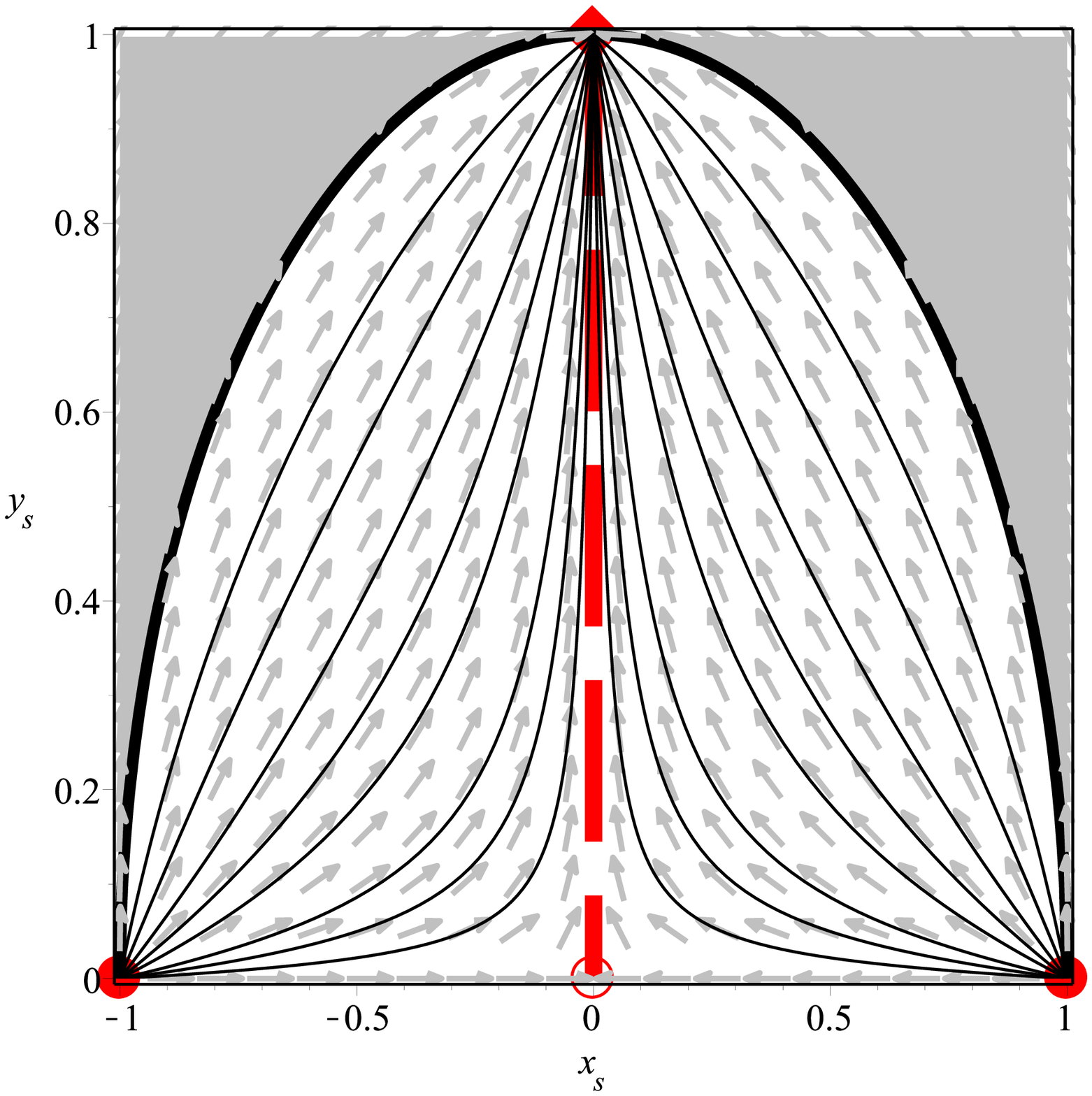}
\includegraphics[width=5cm]{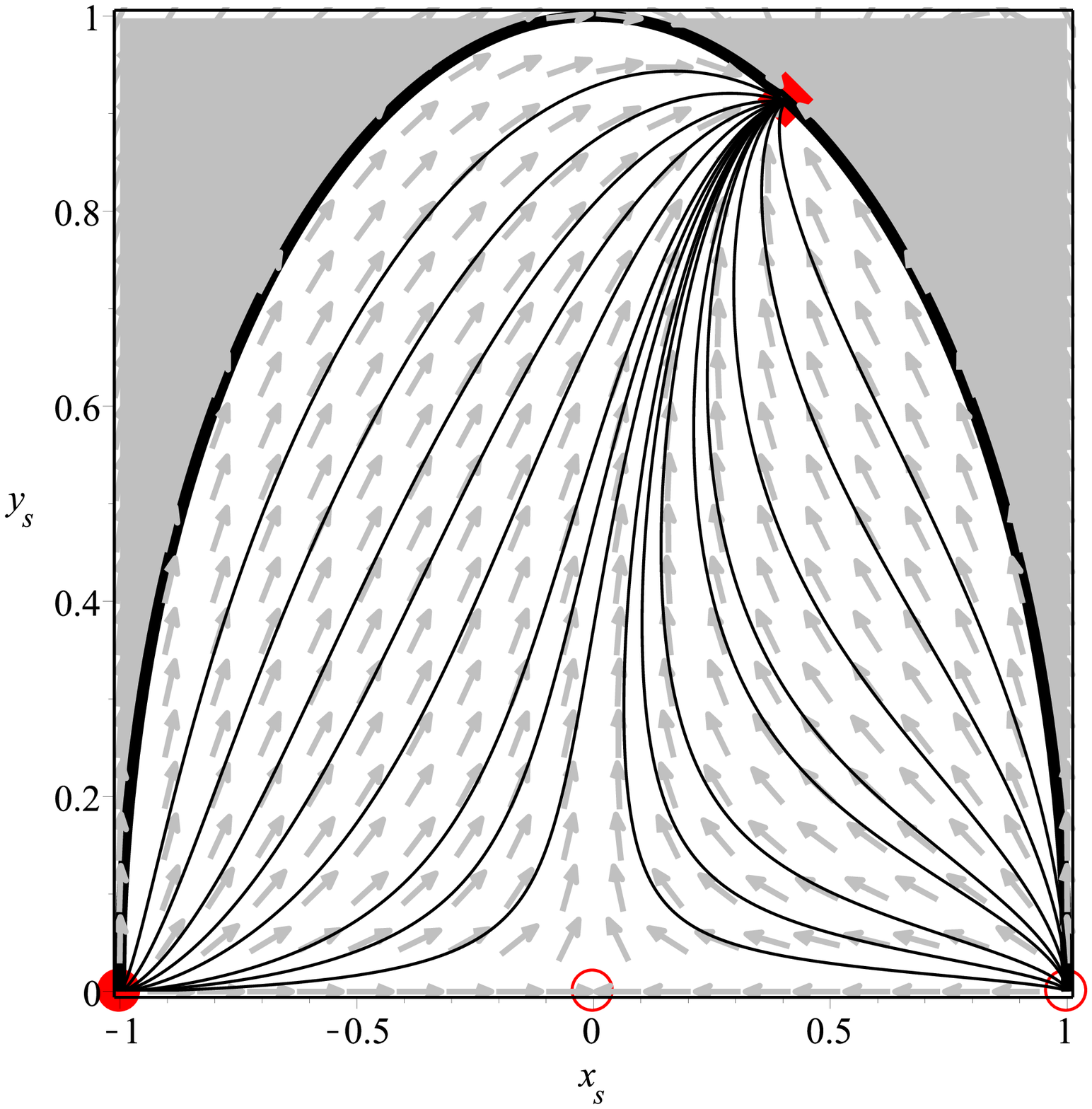}
\includegraphics[width=5cm]{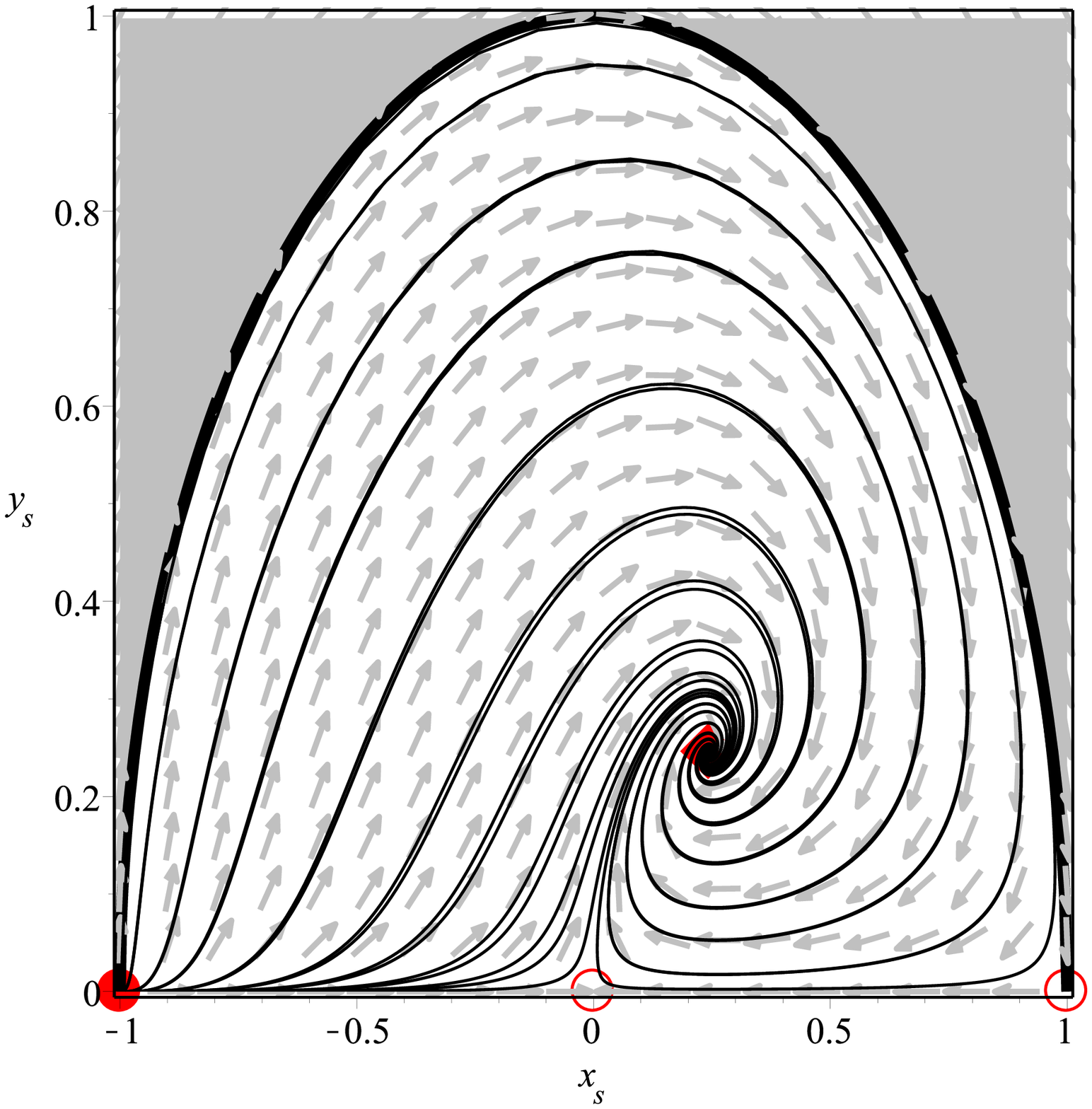}
\includegraphics[width=5cm]{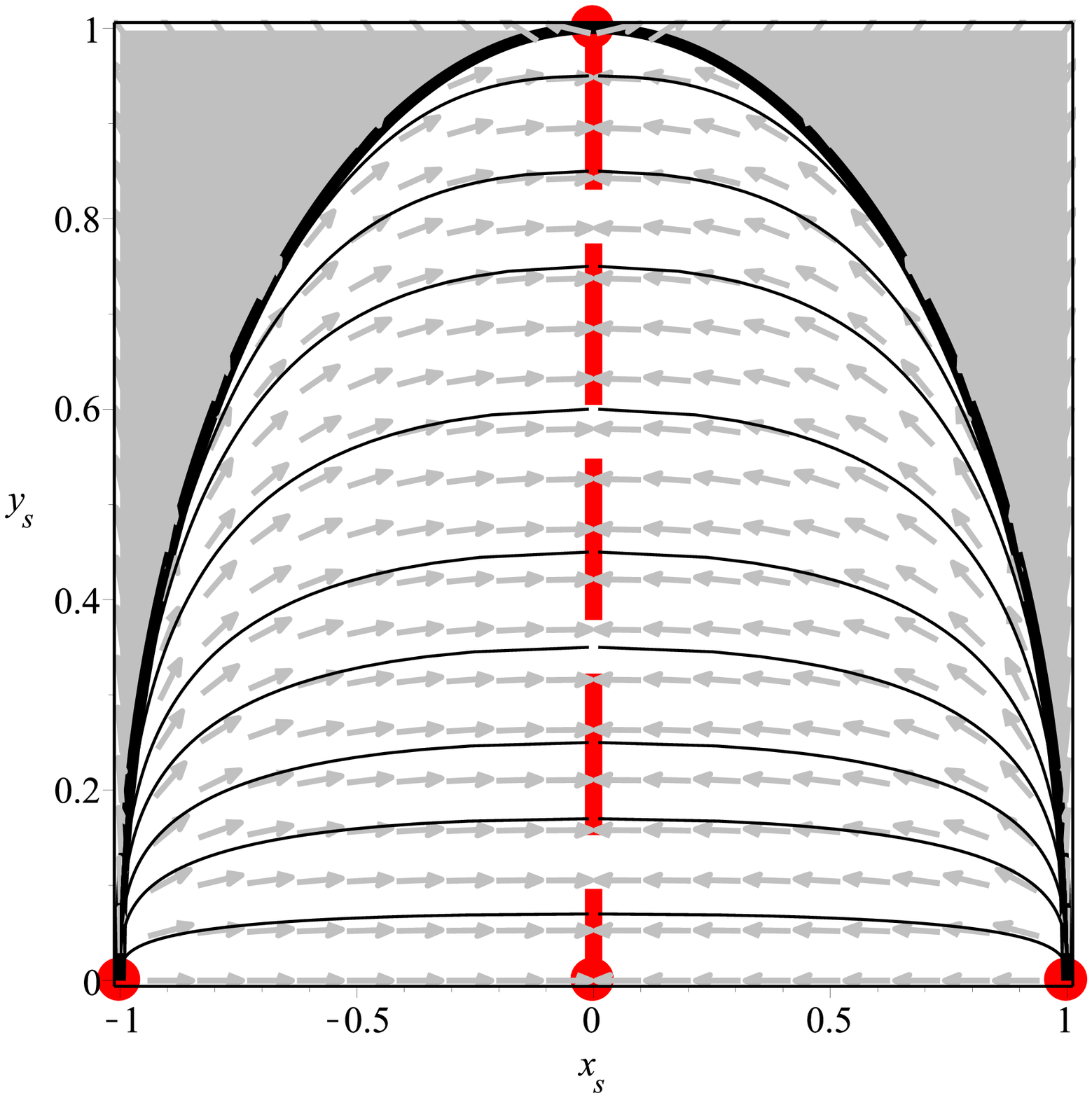}
\includegraphics[width=5cm]{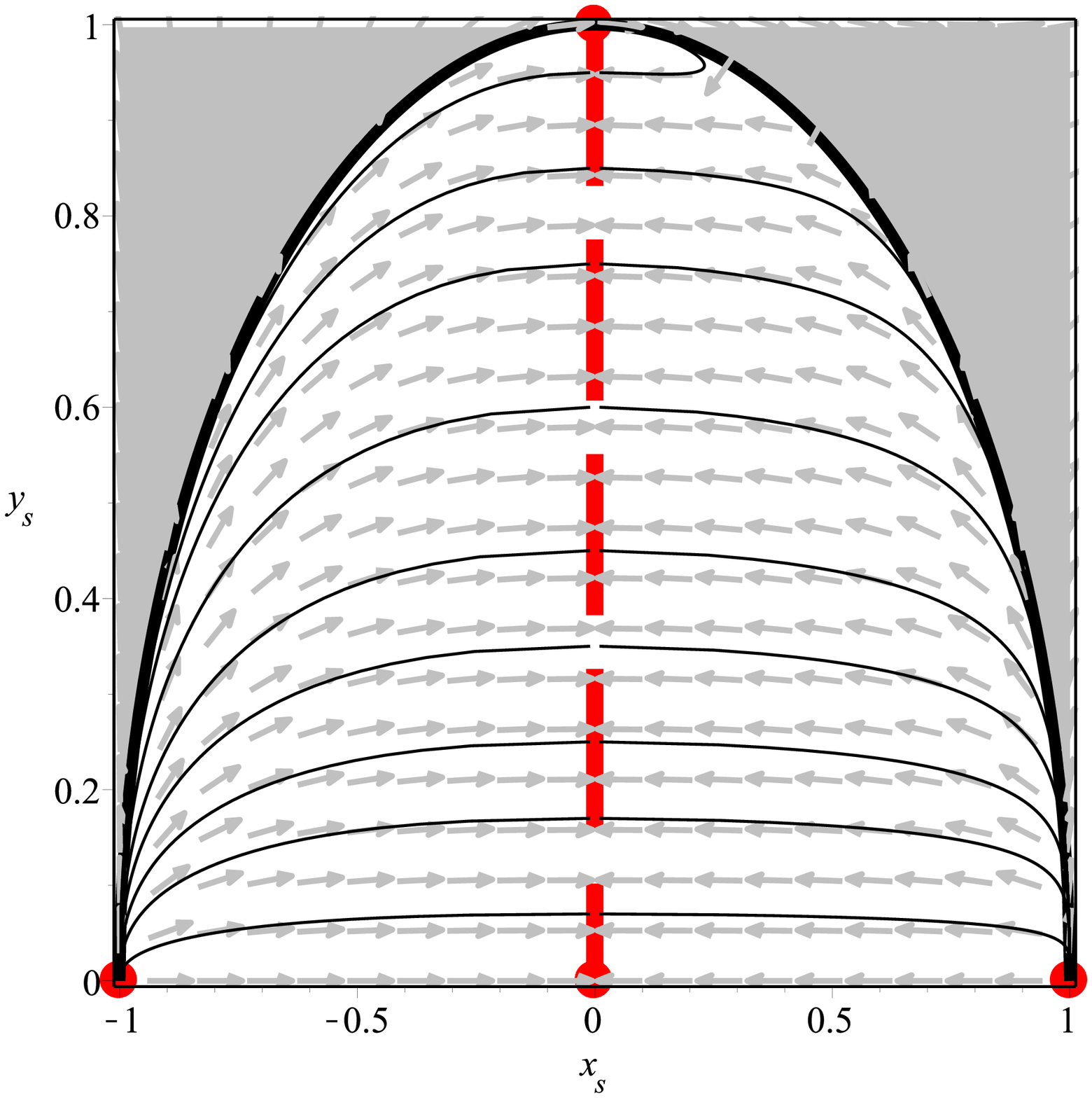}
\includegraphics[width=5cm]{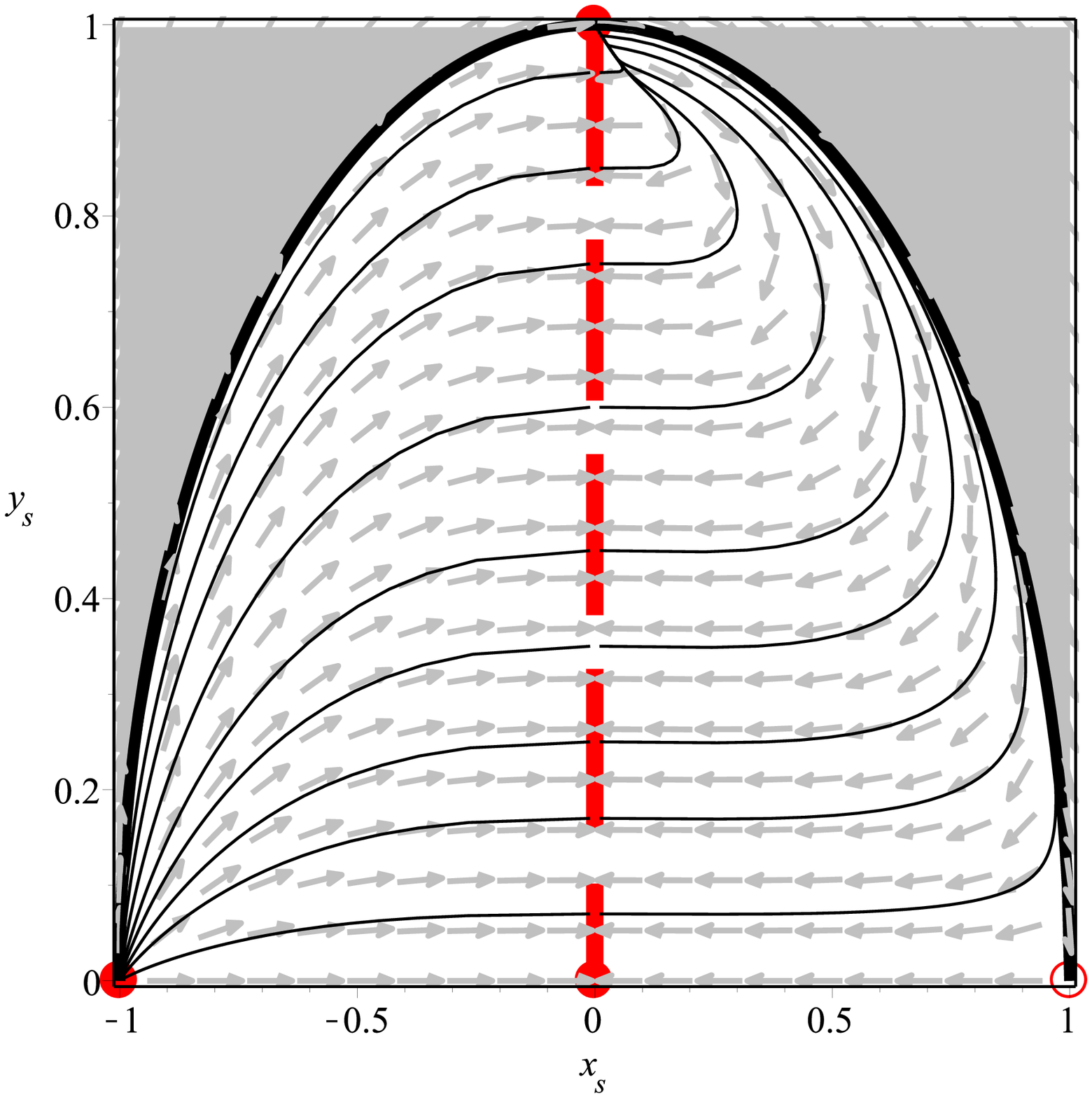}
\includegraphics[width=5cm]{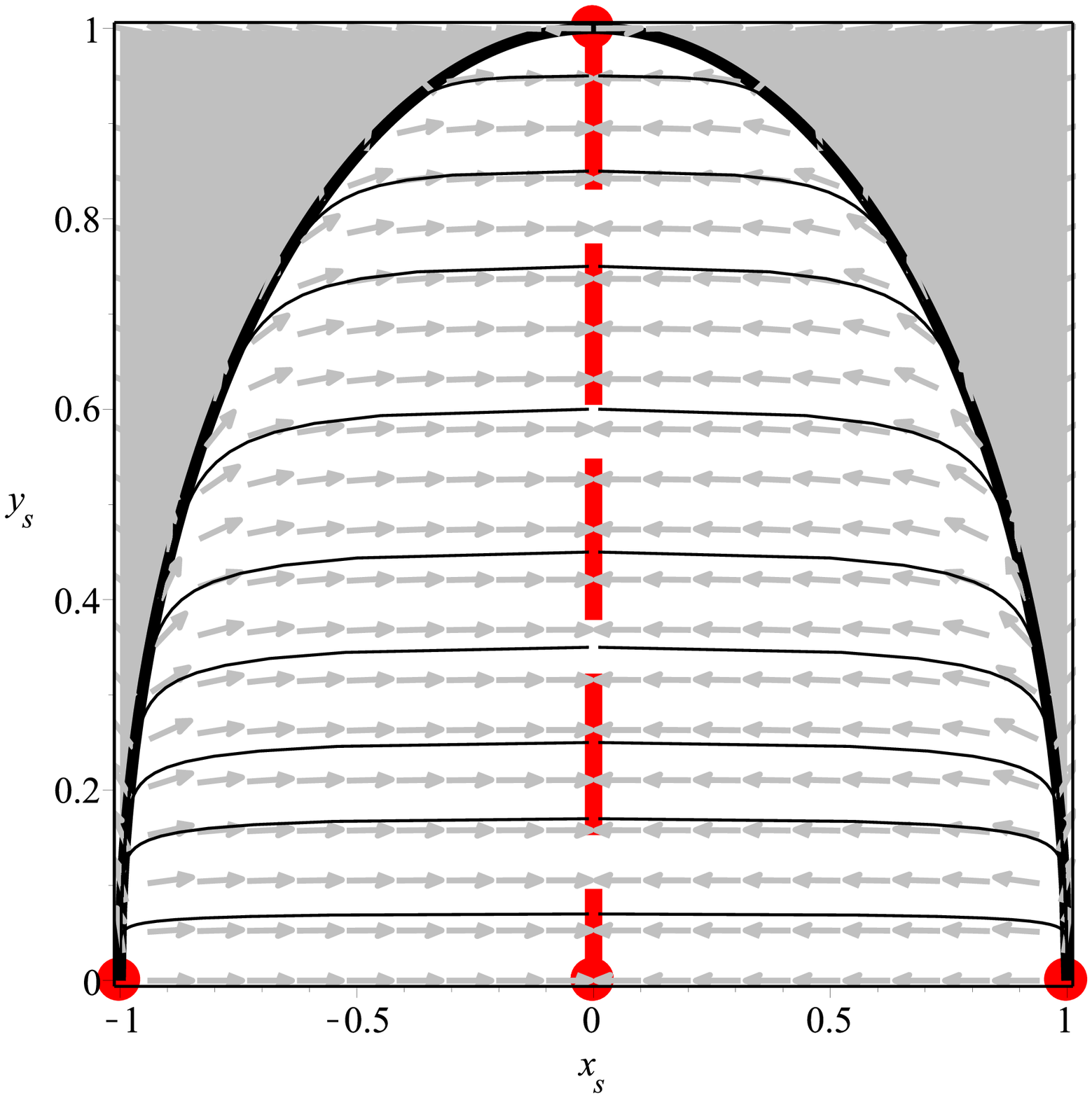}
\includegraphics[width=5cm]{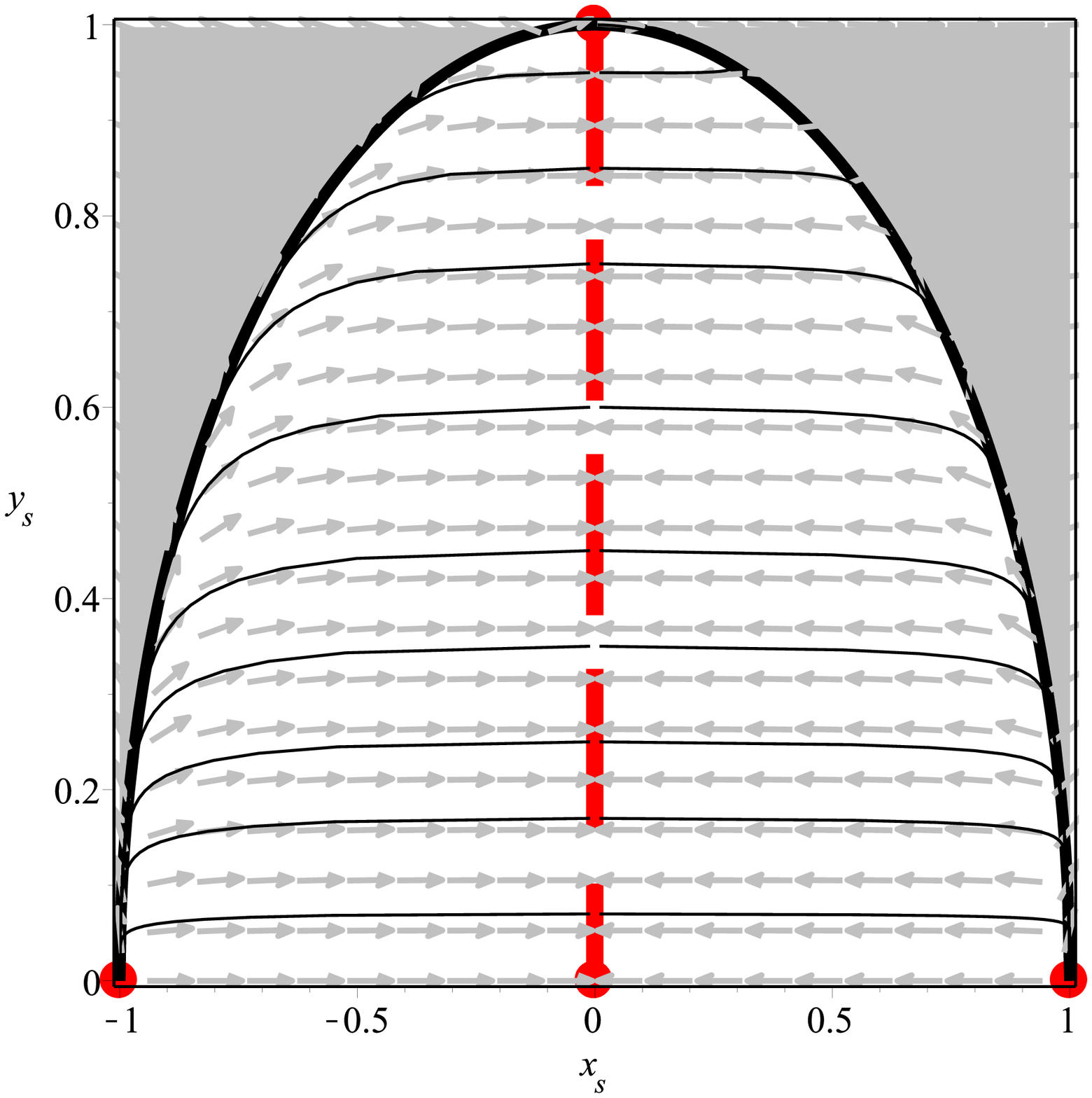}
\includegraphics[width=5cm]{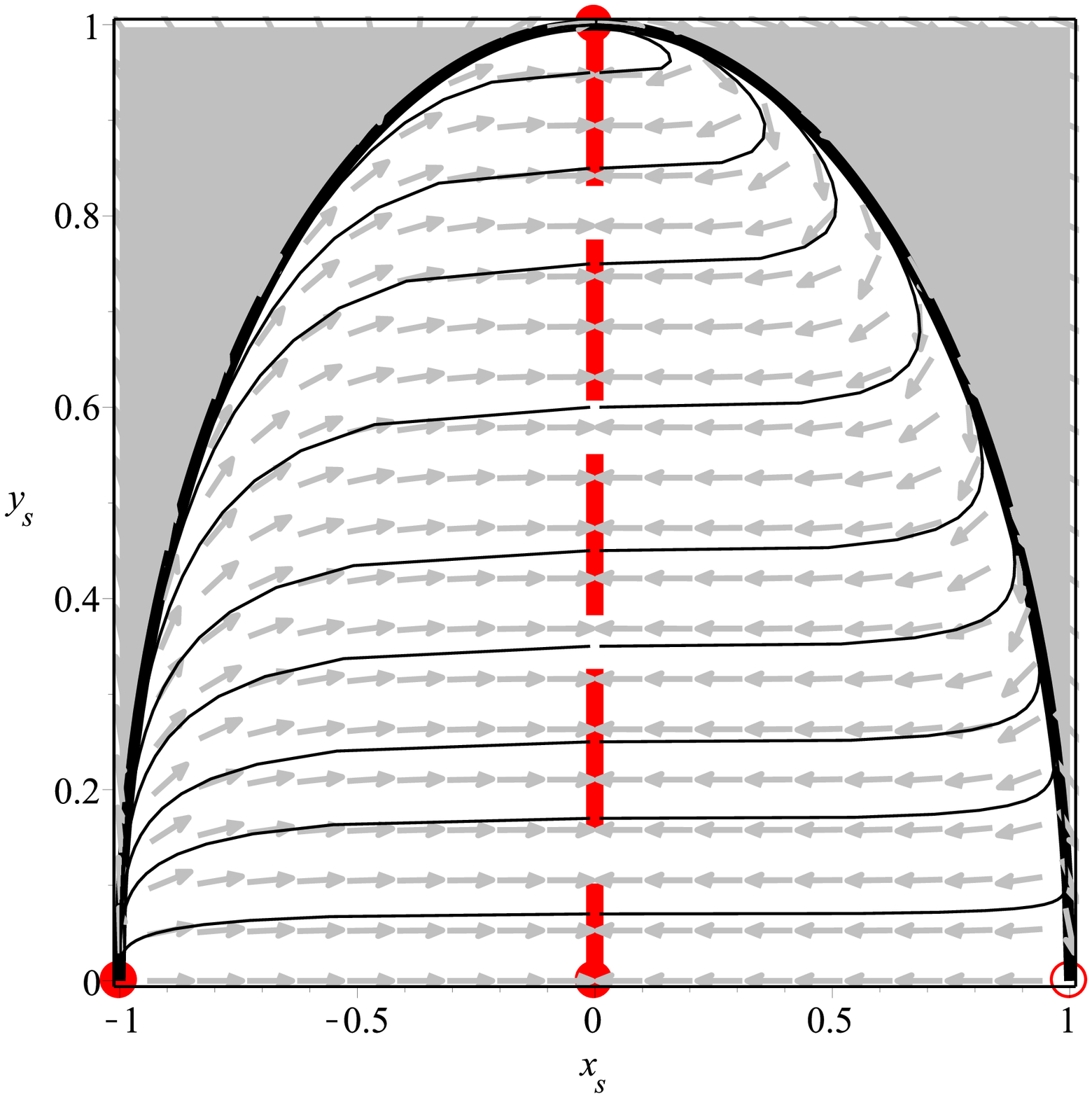}
\caption{Phase portrait for the dynamical system with vanishing coupling to the cubic galileon term $\sigma = 0$, for different choices of the  constants $\alpha$ and $\lambda$. From left to right $\lambda=0,1,5$, while from top to bottom $\alpha=0,1,5$. The figures in the upper row correspond to quintessence without additional non-gravitational interaction with the matter component. The reamining figures are for the interacting quintessence scenario. It is seen that the de Sitter manifold (vertical dash-dot red line at $x_s=0$) is the late-time attractor in this case. For standard quintessence with constant self-interacting potential (left-upper figure) the vertical dash-dot line only represents the separatrix in the phase plane.}\label{figs}\end{figure}



The quintessence model can be retrieved from the galileon model by setting the cubic interacting coupling $\sigma=0$. Actually, if in Eqs. \eqref{ef1}-\eqref{ecc} set $\sigma=0$, one gets the cosmological equations of the interacting quintessence model:

\bea &&3H^2=\rho_m+\rho_\phi,\nonumber\\
&&-2\dot H=(1+\omega_m)\rho_m+\rho_\phi+p_\phi,\nonumber\\
&&\dot\rho_m+3(1+\omega_m)H\rho_m=Q,\nonumber\\
&&\dot\phi\ddot\phi+3H\dot\phi^2=-V_{,\phi}\dot\phi-Q,\label{q-eqs}\eea where the galileon's kinetic and potential energy densities are given by,

\bea \rho_\phi=\frac{\dot\phi^2}{2}+V,\;\;p_\phi=\frac{\dot\phi^2}{2}-V.\nonumber\eea In terms of the phase space variables \eqref{n-var} the quintessence scenario is recovered by setting the variable $z=1$. This means that the dimension of the dynamical system reduces to a plane autonomous system. In this case the Friedmann constraint \eqref{fried-const} simplifies to:

\bea \Omega_m=1-\left(\frac{1\mp x_\pm}{x_\pm}\right)^2-\left(\frac{1-y}{y}\right)^2.\label{q-fried-const}\eea In terms of the standard variables $x_s$ and $y_s$ in Eq. \eqref{normal-vars}, the above constraint takes the much more simple form,

\bea \Omega_m=1-x_s^2-y_s^2.\label{f-const}\eea This means that the standard variables are already normalised and bounded variables: $-1\leq x_s\leq 1$, $0\leq y_s\leq 1$, so that it is not necessary to use the variables $x_\pm$, $y$.

As already mentioned, in the present case the dynamical system corresponding to the cosmological equations \eqref{q-eqs} is a (two-dimensional) plane-autonomous system:

\bea &&x'_s=\frac{\ddot\phi}{\sqrt{6}H^2}-x_s\left(\frac{\dot H}{H^2}\right),\nonumber\\
&&y'_s=-y_s\left(\sqrt\frac{3}{2}\lambda x_s+\frac{\dot H}{H^2}\right),\label{2d-asode}\eea where, as before, the tilde denotes derivative with respect to the number of e-foldings $N=\ln a$, and

\bea &&\frac{\dot H}{H^2}=-\frac{3}{2}\left(1+x_s^2-y_s^2\right),\nonumber\\
&&\frac{\ddot\phi}{H^2}=-3\sqrt{6}x_s+3\lambda y_s^2-\frac{1}{x_s}\left(\frac{Q}{H^3}\right).\label{2d-usef}\eea


The critical points $(x_s,y_s)$ of the dynamical system \eqref{2d-asode} are located within the upper semi-disk $x^2_s+y^2_s\leq 1$ ($y_s\geq 0$). We summarise their properties in Table \ref{tab-7}.

\begin{table*}\centering
\begin{tabular}{||c||c|c|c|c|c|c||}
\hline\hline
Point        & $x_s$   &   $y_s$   &    Stability & $\Omega_m$ & $\omega_\phi$     & Existence \\
\hline\hline
$A^*$    & $1$   &   0  & stable if  &   0  &   1     & always   \\
&&& $\alpha<-1,\;\lambda>\sqrt{6}$ &&&\\
&&& saddle otherwise &&&\\
\hline
$B^*$     & $-1$   &   0  & saddle   &   0  &   1     & always   \\
\hline
$C^*$    & $\frac{\lambda}{\sqrt{6}}$ &  $\sqrt{1-\lambda^2/6}$  & stable if  &   0  &   $\lambda^2/3-1$     &   $ \lambda\leq \sqrt{6}$  \\
&&& $\alpha<1-\lambda^2/3$ &&&\\
&&& saddle otherwise &&&\\
\hline
$D^*$   & $\sqrt{\frac{3}{2}}(\frac{1-\alpha}{\lambda})$   &  $\sqrt{\frac{3(\alpha-1)^2+2\alpha \lambda^2}{2 \lambda^2}}$  &   stable see in Fig. \ref{Fig1}  &   $\frac{(\alpha-1)(\lambda^2+3\alpha-3)}{\lambda^2}$  &   $\frac{- \alpha \lambda^2}{3+\alpha (\lambda^2+3\alpha-6)}$     & $ \alpha \leq 1 \,\, \& \,\, \lambda >0$  \\
&&& saddle otherwise &&&\\
\hline
$E^*$     & $  \sqrt{-\alpha}$   & $0$   &  stable if $\lambda > \frac{1-\alpha}{\sqrt{- 2 \alpha/3}}$       & $1+\alpha$ & $1$ &   $-1 \leq \alpha < 0$ \\
&&& saddle otherwise &&&\\
\hline
$F^*$     & $ - \sqrt{- \alpha}$   & $0$   &    saddle       & $1\!+\!\alpha $ & $1$ &   $-1 \leq \alpha < 0$ \\
\hline
\hline
\end{tabular}\caption{Existence and stability of the critical points as well as several cosmological parameters for interacting quintessence with coupling function $Q=3\alpha H\rho_m$.}\label{tab-7}\end{table*}

\begin{enumerate}

\item The stiff matter points $A^*$ and $B^*$ correspond to solutions dominated by the kinetic term of the scalar field with $\Omega_m =0$. Here the universe is totally governed by the kinetic energy of the quintessence field with EoS $\omega_\phi=1$. These points exist for all values of $\alpha$ and $\lambda$. The positive branch is stable for $\lambda>\sqrt{6}$ and $\alpha<-1$, and saddle otherwise. The negative branch is always  saddle. These solutions correspond to the point $P_6^\pm$ in the cubic galileon case.

\item The scalar field dominated solution $C^*$ implies $\Omega_m=0$ and $x_s^2 + y_s^2 = 1$, meaning that point $C^*$ will always lie on the unit circle.  It exists for $\lambda^2<6$ and it is a stable attractor for $\alpha < 1 -\lambda^2/3$ and a saddle otherwise. This equilibrium point is associated with accelerating expansion of the universe for $\lambda^2<2$. It corresponds to the critical point $P_8$ in the cubic galileon model.

\item The solution $D^*$ is affected by the non-minimal coupling $\alpha$. We have that $\Omega_m\propto\Omega_\phi$, so that it is a scaling solution. The EoS reads  $\omega_\phi=-\alpha\lambda^2/(\alpha(\lambda^2-6)-3(\alpha^2-1))$. The solution depicts an accelerated expansion universe for $\alpha>1/3$. It is stable in the region shown in Fig. \ref{Fig1}, otherwise it is saddle. This point corresponds to $P_9$ in the galileon model with non-minimal interaction and the stability properties are not modified.

\item Solutions $E^*$ and $F^*$ exist only for non-positive coupling $\alpha$ with equation of state parameter $\omega_\phi=1$. The point $E^*$ corresponds to a stable scaling solution for $\lambda>\frac{1-\alpha}{\sqrt{-2\alpha/3}}$, otherwise it is saddle. The point $F^*$ is always a saddle. The points $E^*$ and $F^*$ correspond to the points $P_7^\pm$ in the galileon model.

\end{enumerate}

The above described dynamical behaviour of the interacting quintessence model is illustrated in FIG. \ref{figs} for several values of the free parameters. It is seen that, for the non-interacting quintessence (figures in the upper row in FIG. \ref{figs}), the stiff fluid solutions can be either local past attractors or one of them is the global past attractor while the other one is a saddle point. Hence, in this case the stiff matter is always the origin of the cosmic expansion. This dynamical behaviour is slightly modified in the interacting quintessence scenario where the stiff matter solutions can be either local past attractors or one stiff solution is a local past attractor while the other one is a saddle. In this last case the stiff matter local attractor coexists with a local de Sitter past attractor.

The lifting of the quintessence to galileon by the addition of the cubic interaction quantified by the coupling $\sigma$, enormously improves the resulting cosmological model since the bigbang singularity is always the past attractor. The resulting cosmological scenario is consistent with the present cosmological data and with the standard cosmological model of the universe. This is in contrast to the interacting quintessence model where the past attractor can be either a stiff matter or a de Sitter solution. In the interacting galileon model there are also transient accelerated stages of the early universe that could be associated with early time inflation.


\section{Conclusion}\label{sec-conclu}

In the present paper we have performed a dynamical systems analysis of a cubic galileon with exponential potential $V(\phi)=V_0 e^{-\lambda \phi}$, interacting with pressureless dark matter. We have considered three different coupling functions which are of cosmological interest. Namely $Q_1=3\alpha H \rho_m$,  $Q_2=3\beta\rho_m\dot{\phi}$ and $Q_3 = 3\epsilon H\rho_\phi$, respectively. Applying the dynamical system theory we were able to find the critical points of the autonomous system of equations. For each of these solutions we computed the value of several cosmological parameters of relevance for the physical discussion. We have been able to determine the existence and  stability properties of the critical points. These results are summarized in TABs. \ref{tab-1}-\ref{tab-6} and illustrated in FIGs. \ref{Figq1}-\ref{Figq3}. The most salient cosmological implications of our results are briefly mentioned and discussed below.

The most interesting consequences of the current scenario are the existence of additional critical points besides the solutions of the non-interacting galileon and of the interacting quintessence previously found in the literature. All new critical points correspond to asympotic solutions corresponding to both early time and late time dynamics. We also found slight modifications on the stability properties of each solution as well as of the values of several relevant cosmological parameters. Let us briefly summarise the main modification of the cosmic dynamics due the additional non-gravitational interaction.

\begin{itemize}

    \item The critical points $P_1^\pm$, $R_1^\pm$ and $S_1^\pm$, corresponding to the three different coupling functions considered in this paper, always behave as unstable nodes corresponding to the origin of the space phase trajectories. These are equivalent to the solution $P_1^*$ found in the case of the non-interacting cubic galileon where this critical point is a past attractor. In consequence the singular bigbang is the origin of the cosmic history of our universe. This solution has no analogue in the interacting quintessence scenario.
    
    \item The dark matter dominated solution represents a decelerated phase of the cosmic expansion. The solution behaves always as a saddle equilibrium point. This means that the observed amount of cosmic structure can be accommodated in this model of the universe. It corresponds to the solution $P_7^\pm $ in the limit   $\alpha\rightarrow 0$, and to the solution $R_8^\pm$ in the limit $\beta\rightarrow 0$. These critical points transform to $P_2^*$, $E^*$ and $K^*$ in the corresponding limits. 
    
    \item The de Sitter accelerated expansion, points $P_5^\pm $ and $R_6^\pm$ exists independently of the interacting function. This solution is equivalent to $P_4^*$, $C^*$ and $I^*$ in the limit $\lambda \rightarrow 0$.
    
    \item The related solutions $P_7^\pm$ and $R_8$ have no analogues in the non-interacting cubic galileon scenario. These are equivalent to the critical points $E^*$ and $F^*$, respectively.
    
    \item The galileon dominated solutions $P_8$ and $R_9$, exist both in the non-interacting as well as in the interacting galileon scenarios. 
    
    \item The scaling solutions $P_9$, $R_{10}$ and $S_5$ exist both in the non-interacting cubic galileon as well as in the interacting quintessence scenario. This is due to the fact that the cubic galileon dominates the early time dynamics.
    
    \item Early time inflation solutions $P_2^\pm,$ $R_2^\pm$ and $S_2^\pm$, are saddle critical points in the equivalent phase space, so that these represent a transient stage of the cosmic evolution. In this case the galileon behaves as a cosmological constant. 

    \item The saddle equilibrium points $P_3$ and $S_3$ are associated with the galileon behaving as dust. The solution always exists and it depicts accelerated expansion if $\alpha>2/3$ and $\epsilon>2/3$, respectively. 

    \item The late-time solutions $P_{11}$, $R_{12}$ and $S_9$ are either saddle or isolated local attractors, depending on the free parameters $\lambda$, $\alpha$, $\beta$, $\epsilon$ and on the initial conditions. The galileon EoS $\omega_{\phi}=1$, so that it is a stiff matter solution. This solution is not of interest for the accepted cosmological paradigm. 

\end{itemize}

The main impact of the cubic galileon is to accommodate the bigbang singularity which is a requirement of the accepted cosmological paradigm. The additional non-gravitational interaction between the cubic galileon and the dark matter provides several matter scaling solutions, of which some are late time attractors, thus providing a solution of the cosmic coincidence problem. Besides, as already commented in the introductory part of this paper, generalized cosmological scenarios in which the dark matter and the dark energy interact with each other in a non-gravitational way, can play
an effective role to alleviate/solve the Hubble constant tension \cite{DiValentino2021}.


\section*{Acknowledgments}

The  authors are grateful to SNI-CONACyT for continuous support of their research activity and to FORDECYT-PRONACES-CONACYT for support of the present research under grant CF-MG-2558591 and CF-140630-UNAM-UMSNH.  UN  thanks  the Programa para  el  Desarrollo  Profesional  Docente  of  the  Secretar\'ia de Educaci\'on P\'ublica (PRODEP-SEP) of M\'exico and the Coordinaci\'on  de  la  Investigaci\'on  Cient\'ifica  of  the  Universidad  Michoacana  de  San  Nicol\'as  de  Hidalgo  (CIC-UMSNH) for financial support of his contribution to the present research




\end{document}